\documentstyle[aps,amsmath,amssymb]{revtex}

\def\K{{\cal K}}

\def\M{{\cal M}}
\def\N{{\cal N}}

\def\SS{{\cal S}}

\def\linebreak{\hfill\break}

\def\bra<#1|{\langle #1\rvert}
\def\ket|#1>{\lvert#1 \rangle}
\def\braket<#1|#2>{\langle #1|#2 \rangle}

\def\para{{\scriptscriptstyle /\!/}}
\def\orth{\perp}

\def\tend{\rightarrow}
\def\then{\Rightarrow\quad}

\def\therefore{\mbox{\setbox0=\hbox{X}\hbox{$\ldotp$}\raise0.7\ht0\hbox{$\ldotp$}\hbox{$\ldotp$}} \quad }
\def\because{\mbox{\setbox0=\hbox{X}\raise0.7\ht0\hbox{$\ldotp$}\hbox{$\ldotp$}\raise0.7\ht0\hbox{$\ldotp$}}\kern0pt }

\def\r#1{{\rm #1}}

\def\ZR{{{\mathbb Z}}}

\def\RF{{{\mathbb R}}}

\def\In{\mathrel{\mbox{\setbox0=\hbox{$\cup$}\dimen0=\wd0\divide\dimen0 by 2
\box0\kern -\dimen0\vrule}}}

\def\Frac(#1/#2){\left(\frac{#1}{#2}\right)}

\def\Tdot#1{{{#1}^{\hbox{.}}}}
\def\Tddot#1{{{#1}^{\hbox{..}}}}

\def\Lie{\hbox{\rlap{$\cal L$}$-$}}

\def\Eq#1{\begin{equation} #1 \end{equation}}
\def\Eqr#1{\begin{eqnarray} #1 \end{eqnarray}}

\def\THB{{\mathbb T}}
\def\VHB{{\mathbb V}}
\def\SHB{{\mathbb S}}

\tighten

\begin{document}

\draft


\title{ Brane World Cosmology \\
--- Gauge-Invariant Formalism for Perturbation ---}
\author{Hideo Kodama, Akihiro Ishibashi and Osamu Seto}
\address{
Yukawa Institute for Theoretical Physics, Kyoto University\\
Kyoto 606-8502, Japan
}

\date{April 24, 2000}

\maketitle


\begin{abstract} 
In the present paper the gauge-invariant formalism is developed for
perturbations of the brane-world model in which our universe is
realized as a boundary of a higher-dimensional spacetime. For the
background model in which the bulk spacetime is $(n+m)$-dimensional
and has the spatial symmetry corresponding to the isometry group of a
$n$-dimensional maximally symmetric space, gauge-invariant equations
are derived for perturbations of the bulk spacetime. Further, for the
case corresponding to the brane-world model in which $m=2$ and the
brane is a boundary invariant under the spatial symmetry in the
unperturbed background, relations between the gauge-invariant
variables describing the bulk perturbations and those for brane
perturbations are derived from Israel's junction condition under the
assumption of $\ZR_2$ symmetry. In particular, for the case in which
the bulk spacetime is a constant-curvature spacetime, it is shown that
the bulk perturbation equations reduce to a single hyperbolic master
equation for a master variable, and that the physical condition on the
gauge-invariant variable describing the intrinsic stress perturbation
of the brane yields a boundary condition for the master equation
through the junction condition. On the basis of this formalism, it is
pointed out that it seems to be difficult to suppress brane
perturbations corresponding to massive excitations for a brane motion
giving a realistic expanding universe model.
\end{abstract}



\section{Introduction}

Motivated by the 
M-theory\cite{Horava.P&Witten1996,Horava.P&Witten1996a}, AdS-CFT 
correspondence in string 
theories\cite{Maldacena.J1998,Aharony.O&&1999A}, and the hierarchy 
problem in the particle 
theory\cite{Arkani-Hamed.N1998,Antoniadis.I1998,Randall.L&Sundrum1998A,%
Randall.L&Sundrum1999A}, 
brane-world models in which our universe is realized as a boundary 
of a higher-dimensional spacetime have been actively studied 
recently%
\cite{Randall.L&Sundrum1999Aa,Shiromizu.T&Maeda&Sasaki1999A,%
Chamblin.A&Gibbons1999A,Garriga.J&Tanaka1999A,Tanaka.T&Montes2000A,%
Sasaki.M&Shiromizu&Maeda1999A,Giddings.S&Katz&Randall2000A,%
Chamblin.A&Hawking&Reall1999A,Emparan.R&Horowitz&Myers1999A,%
Emparan.R&Horowitz&Myers1999Aa,Nihei.T1999,Kaloper.N1999,%
Kim.H&Kim1999A,%
Csaki.C&Graesser&Kolda&Terning1999A,Csaki.C&Graesser&Randall&Terning1999B,%
Cline.J&Grojean&Servant1999,%
Flanagan.E&Tye&Wasserman1999A,Binetruy.P&&1999A,Mukohyama.S2000,%
Vollick.D1999A,Kraus.P1999,Ida.D1999A,%
Goldberger.W.D&Wise.M.B1999,%
DeWolfe.O&Freedman.D.Z&Gubser.S.S&Karch.A1999,%
Mukohyama.S&Shiromizu&Maeda1999A,Birmingham.D1999,%
Garriga.J&Sasaki1999A,Cohen.A.G&Kaplan.D.B1999,Gregory.R1999,%
Koyama.K&Soda2000A}. 
In particular, for the case in which the bulk spacetime is 
5-dimensional Anti-de Sitter spacetime and the brane is realized as 
a flat 4-dimensional spacetime, the gravitational interaction 
between matters in the brane is well-described by the standard one 
on scales much larger than the scale corresponding to the brane 
tension%
\cite{Garriga.J&Tanaka1999A,Tanaka.T&Montes2000A,%
Sasaki.M&Shiromizu&Maeda1999A,Giddings.S&Katz&Randall2000A}. 

Further, as an extension of the analysis to a dynamical situation, the
embedding of Robertson-Walker universe models into 5-dimensional
Anti-de Sitter and Anti-de Sitter-Schwarzschild spacetimes has been 
discussed by many people%
\cite{Nihei.T1999,Kaloper.N1999,Kim.H&Kim1999A,%
Csaki.C&Graesser&Kolda&Terning1999A,Csaki.C&Graesser&Randall&Terning1999B,%
Cline.J&Grojean&Servant1999,Flanagan.E&Tye&Wasserman1999A,%
Binetruy.P&&1999A,Mukohyama.S2000,Vollick.D1999A,Kraus.P1999,%
Ida.D1999A}. 
In such high-symmetry cases, although the evolution equation 
for the cosmic scale factor is modified from the standard one, our 
universe is still a dynamically closed system, and the difference 
in the evolution  equation can be neglected when the energy density 
of the universe becomes much smaller than the brane tension. Thus 
the brane-world model gives a new world model consistent with the 
present day observations. However, if one goes beyond this 
lowest-level approximation, it is not clear whether the brane-world 
model is consistent with all available observations because our 
universe is not dynamically closed in this model\cite{Shiromizu.T&Maeda&Sasaki1999A}. 

One of the simplest ways to analyze this problem is to investigate the
behavior of perturbations of the brane-world model. Since
perturbations of the brane are inevitably associated with
perturbations in the geometry of the bulk spacetime, such
investigation will make clear whether or not the open nature of the universe
dynamics is controllable. It will also make possible an observational 
test of the model in terms of the anisotropy of the
cosmic microwave background.

As the starting point of investigations in this line, in the present
paper, we develop a gauge-invariant formalism for perturbations of the
brane-world model. The basic approach is the same as that originally
developed for 4-dimensional spacetime by Gerlach and Sengupta%
\cite{Gerlach.U&Sengupta1979,Gerlach.U&Sengupta1979a,Gerlach.U&Sengupta1979b} 
and utilized  by some people in the analysis of the 
interaction between a domain wall and gravitational waves in 
4-dimensional spacetimes%
\cite{Kodama.H&Ishihara&Fujiwara1994,Ishibashi.A&Ishihara1997,Ishibashi.A&Ishihara1999}. 

The formalism consists of two parts. The first one is a
gauge-invariant formalism for perturbations in the geometry of the
bulk spacetime. This problem has already been investigated by some
people for the standard case in which the bulk spacetime is vacuum and
maximally symmetric\cite{Mukohyama.S2000A}. In the present paper,
taking account of the developing nature of the brane-world model, we
extend the formalism to the case in which the bulk spacetime is
$(m+n)$-dimensional and its unperturbed geometry has only the isometry
corresponding to the maximally symmetric space of dimension
$n$($n\ge1$). This symmetry is utilized to expand perturbations in
terms of the harmonic functions on $n$-dimensional maximally symmetric
space and define gauge-invariant variables.

The second part establishes relations between the gauge-invariant
variables describing perturbations of the brane and those for the bulk
perturbations. In this part we assume that $m=2$ and the
$(n+1)$-dimensional brane is invariant under the isometry group of the
bulk in the unperturbed model. Thus the brane represents an expanding
Robertson-Walker universe in general.

The paper is organized as follows. In the next section we first
classify perturbations into tensor, vector, and scalar types in terms
of the tensorial behavior with respect to the maximally symmetric
$n$-dimensional spacetime. Then for each type we define the
gauge-invariant variables describing perturbations of the bulk
geometry and express the Einstein equations in terms of them. In
Section III, after introducing an gauge-invariant variable describing
the motion of the brane, we express Israel's junction condition
corresponding to the $\ZR_2$ symmetry in terms of it and the bulk
variables. We will show that this gives expressions for the intrinsic
perturbation variables for the brane in terms of the bulk variables, 
and a boundary condition on the latter in terms of the intrinsic
stress perturbations of the brane. In Section IV we specialize the
formalism to the standard brane-world model in which the bulk
spacetime is vacuum. We reduce the perturbation equations to a single
hyperbolic equation for a master variable $\Omega$ in a two
dimensional spacetime and express the junction conditions in terms of
the master variable. We will show that the condition that the
anisotropic stress perturbation of the brane should vanish yields the
Neumann and Dirichlet boundary condition on the master variable
for the tensor and vector perturbations, respectively, while the
boundary condition for the scalar perturbation is obtained from the
condition on the entropy perturbation of the brane. The last condition 
becomes non-local with respect to time except for the cases
in which the brane is vacuum or $p=-\rho$. Section V is devoted to
summary and discussion.


\section{Bulk Perturbation Equations}
\label{sec:BulkPerturbationEq}


\subsection{Background spacetime}
\label{sec:BulkBackground}

In this section we consider perturbations of spacetime structure on
$(m+n)$-dimensional spacetime $\M$ which is locally written as a product
\Eq{
\M^{m+n}=\N^m\times \K^n \ni (y^a, x^i)=(z^M).
}
Its unperturbed background geometry is given by the metric
\Eq{
d{\bar s}^2={\bar g}_{MN}dz^M dz^N
=g_{ab}(y)dy^a dy^b + r^2(y)d\sigma_n^2,
}
where the metric
\Eq{
d\sigma_n^2=\gamma_{ij}(x)dx^i dx^j 
}
is that with a constant sectional curvature $K$ on $\K^n$.  We denote
the covariant derivatives, the connection coefficients, 
and the curvature tensors for the three metrics $d{\bar s}^2$, 
$g_{ab}dy^a dy^b$, and $d\sigma_n^2$ as
\Eqr{
& d{\bar s}^2 & \then 
{\bar \nabla}_M, {\bar \Gamma}^M_{NL}, {\bar R}_{MNLS}, \\
& g_{ab}(y)dy^a dy^b & \then 
D_a, {}^m\!\Gamma^a_{bc}(y), {}^m\! 
R_{abcd}(y), \\
& d\sigma_n^2 & \then \hat D_i, \hat\Gamma^i_{jk}(x),
\hat R_{ijkl}(x)=K(\gamma_{ik}\gamma_{jl}-\gamma_{il}\gamma_{jk}).
}
The expressions for ${\bar \Gamma}^M_{NL}$ and ${\bar R}_{MNLS}$ in 
terms of the corresponding quantities for the metric $g_{ab}(y)dy^a 
dy^b$ and $d\sigma_n^2$ are given in Appendix \ref{Appendix:A}.

From the symmetry structure of ${\bar G}_{MN}$ the energy-momentum 
tensor ${\bar T}_{MN}$ for the background bulk geometry has the 
structure 
\Eq{
{\bar T}_{ai}=0,\ {\bar T}^i_j={\bar P}\delta^i_j.
}
Hence the Einstein equations for the bulk spacetime
\Eq{
{\bar G}_{MN}+\Lambda {\bar g}_{MN}=\kappa^2 {\bar T}_{MN}
}
are reduced in the unperturbed background to
\Eqr{
& {\bar G}_{ab}+\Lambda g_{ab}=\kappa^2 {\bar T}_{ab},\\
& {\bar G}^i_i=n(\kappa^2 {\bar P}-\Lambda).
}
%


\subsection{Gauge-transformation of perturbations}

For the infinitesimal gauge transformation represented in terms of the
coordinates as $\bar\delta z^M=\xi^M$, the metric
perturbation $h_{MN}=\delta {\bar g}_{MN}$ transforms as
\Eq{
\bar\delta h_{MN}=-\Lie_\xi g_{MN}
=-\bar\nabla_M\xi_N-\bar\nabla_N\xi_M.
}
By decomposing the connection this yields
\Eqr{
&& \bar\delta h_{ab}=-D_a\xi_b-D_b\xi_a, 
\label{GaugeTrf:metric:S}\\
&& \bar\delta 
h_{ai}=-r^2D_a\left(\frac{\xi_i}{r^2}\right)
-\hat D_i\xi_a,
\label{GaugeTrf:metric:V}\\
&& \bar\delta h_{ij}=-\hat D_i\xi_j-\hat D_j\xi_i-2rD^ar \xi_a 
\gamma_{ij}.
\label{GaugeTrf:metric:T}
}
Similarly, the gauge transformation of the perturbation of the 
energy-momentum tensor $\bar\delta(\delta {\bar T})_{MN}$,
\Eq{
\bar\delta(\delta {\bar T})_{MN}=-\Lie_\xi {\bar 
T}_{MN}=-\xi^L\bar\nabla_L {\bar T}_{MN}-{\bar 
T}_{ML}\bar \nabla_N\xi^L-{\bar T}_{NL}\bar\nabla_M\xi^L
}
is written as
\Eqr{
&& \bar\delta(\delta {\bar T})_{ab}=-\xi^cD_c{\bar T}_{ab}-{\bar 
T}_{ac}D_b\xi^c-{\bar T}_{bc}D_a\xi^c,
\label{GaugeTrf:EM:S}\\
&& \bar\delta(\delta {\bar T})_{ai}=-{\bar T}_{ab}\hat 
D_i\xi^b-r^2{\bar P}D_a(r^{-2}\xi_i),
\label{GaugeTrf:EM:V}\\
&& \bar\delta(\delta {\bar T})_{ij}=-\xi^aD_a(r^2{\bar 
P})\gamma_{ij}-{\bar P}(\hat D_i\xi_j+\hat 
D_j\xi_i).
\label{GaugeTrf:EM:T}}
%


\subsection{Gauge-Invariant Perturbation Equations}

In general, each tensor with rank at most 2 on the maximally symmetric
space $\K^n$ is uniquely decomposed into components of the three types,
scalar, vector, and tensor, and each component can be expanded in terms
of harmonic functions of the same type\cite{Kodama.H&Sasaki1984}.


\subsubsection{Tensor perturbation}
\label{sec:BulkPerturbation:tensor}

First we consider the tensor perturbation, which can be expanded in 
terms of the harmonic tensors $\THB_{ij}$, 
\Eq{
(\hat\triangle + k^2)\THB_{ij}=0,
}
with the properties
\Eq{
\THB^i_i=0, \quad \hat D_j \THB^j_i=0.
}
In the present paper we omit the index labeling the harmonics as well
as the summation symbol with respect to the index, because expansion
coefficients corresponding to different eigenvalues decouple on the
maximally symmetric space.

Here note that the eigenvalue $k^2$ is always non-negative under a
boundary condition making the operator $\hat\triangle$ self-adjoint in
the $L^2$ space. In particular, $k^2=0$ appears only for the flat
space ($K=0$) since the corresponding eigentensors satisfy $\hat
D_k\THB_{ij}=0$, which yields $0=\hat D^i\hat
D_k\THB_{ij}=nK\THB_{jk}$. Thus the eigentensors for $k^2=0$ are
constant tensors. In the framework of the expansion in the $L^2$
sense, such eigentensors should be discarded. Thus we assume $k^2>0$
in the following unless otherwise stated.

For the tensor perturbation the metric perturbation is expanded as
\Eq{
h_{ab}=0,\quad 
h_{ai}=0,\quad
h_{ij}=2r^2 H_T \THB_{ij}.
}
Since the infinitesimal gauge transformation $\xi=(\xi^a,\xi^i)$ has no tensor
component, it follows that $H_T$ is gauge-invariant. Similarly,
$\delta {\bar T}_{MN}$ is expanded as
\Eq{
\delta {\bar T}_{ab}=0,\ 
\delta {\bar T}^a_{i}=0,\ 
\delta {\bar T}^i_{j}=\tau_T \THB^i_{j},
}
where $\tau_T$ is the gauge-invariant variable representing the 
tensor-type anisotropic stress perturbation.

Inserting these expansions into the expression for $\delta {\bar 
R}_{ij}$, we obtain the following gauge-invariant perturbation 
equation:
\Eq{
-\square H_T-\frac{n}{r}Dr\cdot DH_T+ 
\frac{k^2+2K}{r^2}H_T=\kappa^2\tau_T, 
\label{BulkPerturbationEq:tensor}}
where $\square=D^aD_a$ is the d'Alembertian on the $m$-dimensional 
space $\N^m$.


\subsubsection{Vector Perturbation}
\label{sec:BulkPerturbation:vector}

Divergence-free vector fields can be expanded in terms of the vector 
harmonic $\VHB_i$ defined by
\Eqr{
&& (\hat\triangle +k^2)\VHB_i=0,\\
&& \hat D_i \VHB^i=0.
}
From this we can define the vector-type harmonic tensor as
\Eq{
\VHB_{ij}=-\frac{1}{2k}(\hat D_i\VHB_j+\hat D_j\VHB_i),
\label{VHB:tensor:def}}
which has the properties
\Eqr{
&& \left[\hat \triangle +k^2-(n+1)K\right]\VHB_{ij}=0,\\
&& \VHB^i_i=0,\quad
\hat D_j \VHB^j_i=\frac{k^2-(n-1)K}{2k}\VHB_i,
\label{VHB:tensor:div}}
and expands a vector-type perturbation of a 2nd-rank tensor.

As in the case of tensor harmonics, the eigenvalue $k^2$ is always
non-negative and $k^2=0$ occurs only for $K=0$, for which the harmonic
vectors become constant vectors. Thus, for the same reason as in the
tensor harmonics, we assume $k^2>0$ in the following. One subtle point
of the vector harmonics is that $k^2>0$ does not imply 
$k^2-(n+1)K>0$ for $K>0$. Hence for $k^2<(n+1)K$ and $K>0$, the
vector-type tensor harmonics defined by
\eqref{VHB:tensor:def} should vanish, which implies that $\VHB^i$ is 
a Killing vector on $S^n$. In this case it follows from 
\eqref{VHB:tensor:div} that the eigenvalue should be given by 
$k^2=(n-1)K$.

The vector perturbation of the metric is expanded in terms of the 
vector harmonics as
\Eq{
h_{ab}=0,\quad
h_{ai}=rf_a \VHB_i,\quad
h_{ij}=2r^2 H_T \VHB_{ij},
}
and the vector perturbation of the energy-momentum tensor as
\Eq{
\delta {\bar T}_{ab}=0,\ 
\delta {\bar T}^a_{i}=r\tau^a \VHB_i,\ 
\delta {\bar T}^i_{j}= \tau_T \VHB^i_{j}. 
}
For the reason stated above, $H_T$ and $\tau_T$ are not defined for 
the mode $k^2=(n-1)K$ with $K>0$.

Since the infinitesimal gauge transformation $\xi$ has only 
the vector component
\Eq{
    \xi_a=0,\ \xi_i=rL \VHB_i,
}
the expansion coefficients of the perturbation transform as
\Eq{
\bar\delta f_a=-rD_a\Frac(L/r),\
\bar\delta H_T=\frac{k}{r}L,\
\bar\delta \tau_a=0,\
\bar\delta \tau_T=0.
}
Hence, except the mode $k^2=(n-1)K$ for $K>0$, the vector 
perturbation is described by the three 
gauge-invariant variables $\tau_a$, $\tau_T$ and 
\Eq{
F_a=f_a+\frac{r}{k}D_a H_T.
}
On the other hand, for the mode $k^2=(n-1)K$ with $K>0$, only the 
combination
\Eq{
F^{(1)}_{ab}=rD_a\Frac(f_b/r)-rD_b\Frac(f_a/r)
}
is gauge-invariant.

From the components $\delta {\bar G}^a_i$ and $\delta {\bar G}^i_j$ 
of the Einstein equations we obtain the following gauge-invariant 
perturbation equations except the mode $k^2=(n-1)K$ with $K>0$:
\Eqr{
&& \frac{1}{r^{n+1}}D^b\left[r^{n+2}\left\{
D_b\Frac(F_a/r)-D_a\Frac(F_b/r)\right\}\right]
-\frac{k^2-(n-1)K}{r^2}F_a
=-2\kappa^2\tau_a, \label{BulkPerturbationEq:vector1}\\
&& \frac{k}{r^{n}}D_a(r^{n-1}F^a)=-\kappa^2 \tau_T.
\label{BulkPerturbationEq:vector2}}
On the other hand, for the mode $k^2=(n-1)K$ with $K>0$, the second 
equation does not appear and the first equation is written as
\Eq{
\frac{1}{r^{n+1}}D^b\left(r^{n+1}F^{(1)}_{ab}\right)= - 2\kappa^2 \tau_a.
\label{BulkPerturbationEq:vector3}}
%


\subsubsection{Scalar Perturbation}
\label{sec:BulkPerturbation:scalar}

From the scalar harmonic functions
\Eq{
(\hat \triangle + k^2)\SHB=0,
}
we can construct the scalar-type harmonic vectors $\SHB_i$ as
\Eqr{
&& \SHB_i=-\frac{1}{k}\hat D_i \SHB,\\
&& [\hat \triangle + k^2-(n-1)K]\SHB_i=0,\\
&& \hat D_i \SHB^i=k\SHB,
}
and the scalar-type harmonic tensors $\SHB_{ij}$ as
\Eqr{
&& \SHB_{ij}=\frac{1}{k^2}\hat D_i\hat D_j \SHB +\frac{1}{n}\gamma_{ij}\SHB,\\
&& \SHB^i_i=0,\quad
\hat D_j \SHB^j_i=\frac{n-1}{n}\frac{k^2-nK}{k}\SHB_i,\\
&& [\hat \triangle +k^2-2nK]\SHB_{ij}=0.}

In contrast to the vector and tensor harmonics, a constant function
becomes the normalizable $k=0$-mode for $K>0$, for which $\SHB_i$ and
$\SHB_{ij}$ vanish identically. Since $\SHB_i\equiv0$ implies
$\SHB$=const, no degeneracy occurs for the scalar-type harmonic
vectors except for this constant mode, and $k^2>(n-1)K$ if $k^2>0$. On
the other hand, $\SHB_{ij}$ vanishes identically for $k^2=nK$. For
$k^2>0$ this occurs only for $K>0$. Since the spectrum of $k^2$ is
given by $k^2=l(l+n-1)K$ with non-negative integer $l$, it corresponds
to the $l=1$ harmonics. For other modes $k^2>2nK$.

A scalar perturbation of the metric is expanded in terms of the 
scalar harmonics as
\Eq{
h_{ab}=f_{ab}\SHB,\ 
h_{ai}=rf_a \SHB_i,\ 
h_{ij}=2r^2(H_L\gamma_{ij}\SHB+H_T \SHB_{ij}),
}
and a scalar perturbation of the energy-momentum tensor as
\Eq{
\delta {\bar T}_{ab}=\tau_{ab}\SHB,\
\delta {\bar T}^a_i=r\tau^a \SHB_i,\
\delta {\bar T}^i_j=\delta {\bar P} \delta^i_j \SHB + \tau_T 
\SHB^i_j.
}
In these expansions terms corresponding to $H_T$ and $\tau_T$ for 
$k^2=nK>0$ and those corresponding to $f_a,H_T, \tau_a$ and 
$\tau_T$ for $k^2=0$ do not exist.

For $k^2(k^2-nK)\not=0$, under the infinitesimal gauge transformation
\Eq{
    \xi_a=T_a \SHB,\ \xi_i=rL \SHB_i,
}
these expansion coefficients transform as
\Eqr{
&& \bar\delta f_{ab}=-D_a T_b -D_b T_a,\\
&& \bar\delta f_a=-rD_a\Frac(L/r)+\frac{k}{r}T_a,\\
&& \bar\delta X_a=T_a,\\
&& \bar\delta H_L=-\frac{k}{nr}L-\frac{D^ar}{r}T_a,\\
&& \bar\delta H_T=\frac{k}{r}L,\\
&& \bar\delta \tau_{ab}=-T^cD_c{\bar T}_{ab}-{\bar 
T}_{ac}D_bT^c-{\bar T}_{bc}D_aT^c,\\
&& \bar\delta 
\tau_a=\frac{k}{r}({\bar T}_{ab}T^b-{\bar P}T_a),\\
&& \bar\delta 
(\delta {\bar P})=-T^aD_a {\bar P},\\ 
&& \bar\delta \tau_T=0,
}
where $X_a$ is defined as
\Eq{
X_a=\frac{r}{k}\left(f_a+\frac{r}{k}D_a H_T\right).
}
Hence, in addition to $\tau_T$ we can construct 5 independent 
gauge-invariant quantities as
\Eqr{
&& F=H_L+\frac{1}{n}H_T+\frac{1}{r}D^ar X_a,\\
&& F_{ab}=f_{ab}+D_aX_b+D_bX_a,\\
&& \Sigma_{ab}=\tau_{ab}+{\bar T}^c_b D_aX_c+{\bar 
T}^c_aD_bX_c+X^cD_c{\bar T}_{ab},\\
&& \Sigma_a=\tau_a -\frac{k}{r}({\bar T}^b_a X_b - {\bar P}X_a),\\
&& \Sigma = \delta {\bar P} +X^aD_a{\bar P}.
}
On the other hand, for the modes $k^2(k^2-nK)=0$, these become 
gauge-dependent if we define them by putting undefined variables to 
zero.

From the components $\delta {\bar G}_{ab}$, $\delta {\bar G}^a_i$, 
$\delta {\bar G}^i_i $ and the traceless part of $\delta G^i_j$ of 
the Einstein equations, we obtain the following four 
gauge-invariant perturbation equations for modes 
$k^2(k^2-nK)\not=0$:
\Eqr{
&&-\square F_{ab}+D_aD_c F^c_b+D_bD_cF^c_a
+n\frac{D^cr}{r}(-D_cF_{ab}+D_aF_{cb}+D_bF_{ca}) \nonumber\\
&&\quad +\,{}^m\!R^c_aF_{cb}+\,{}^m\!R^c_bF_{ca}
-2\,{}^m\!R_{acbd}F^{cd}+\left(\frac{k^2}{r^2}-{\bar 
R}+2\Lambda\right)F_{ab}\nonumber\\
&&\quad -D_aD_b 
F^c_c-2n\left(D_aD_bF+\frac{1}{r}D_arD_bF+\frac{1}{r}D_brD_aF 
\right) \nonumber\\
&&\quad -\left[ D_cD_dF^{cd}+\frac{2n}{r}D^cr 
D^dF_{cd}\right.\nonumber\\
&&\quad +\left(-{}^m\!R^{cd}+\frac{2n}{r}D^cD^dr
+\frac{n(n-1)}{r^2}D^crD^dr\right)F_{cd} \nonumber\\
&&\quad -2n\square F -\frac{2n(n+1)}{r}Dr\cdot 
DF+2(n-1)\frac{k^2-nK}{r^2}F \nonumber\\
&&\quad  \left. -\square 
F^c_c-\frac{n}{r}Dr\cdot DF^c_c+\frac{k^2}{r^2}F^c_c \right]g_{ab} 
 =2\kappa^2 \Sigma_{ab},
\label{BulkPerturbationEq:scalar1}\\
&& \frac{k}{r}\left[-\frac{1}{r^{n-2}}D_b(r^{n-2}F^b_a)
+rD_a\left(\frac{1}{r}F^b_b\right)+2(n-1)D_aF\right] 
=2\kappa^2\Sigma_a,
\label{BulkPerturbationEq:scalar2}}
\Eqr{
&&-\frac{1}{2}D_aD_b F^{ab}-\frac{n-1}{r}D^arD^bF_{ab} \nonumber\\
&&\quad +\left[\frac{1}{2}{}^m\!R^{ab}-\frac{(n-1)(n-2)}{2r^2}D^arD^br-(n-1)
\frac{D^aD^br}{r}\right]F_{ab} \nonumber\\
&&\quad +\frac{1}{2}\square F^c_c+\frac{n-1}{2r}Dr\cdot DF^c_c
-\frac{n-1}{2n}\frac{k^2}{r^2}F^c_c \nonumber\\
&&\quad +(n-1)\square F + \frac{n(n-1)}{r}Dr\cdot DF
-\frac{(n-1)(n-2)}{n}\frac{k^2-nK}{r^2}F 
=\kappa^2\Sigma,
\label{BulkPerturbationEq:scalar3}\\
&& -\frac{k^2}{2r^2}\left[2(n-2)F+ F^a_a\right]=\kappa^2 \tau_T.
\label{BulkPerturbationEq:scalar4}}

For the exceptional case $k^2=nK>0$
\eqref{BulkPerturbationEq:scalar4} does not exist, and for the case 
$k^2=0$ \eqref{BulkPerturbationEq:scalar2} and 
\eqref{BulkPerturbationEq:scalar4} do not appear. The other 
equations still hold although each variable is gauge-dependent.

Here, note that from the Bianchi identities not all of these equations
are independent, and some combinations of them yield the
energy-momentum conservation law for the bulk matter perturbation. For
example, if we eliminate $D_b F^b_a$ and $F^a_a$ in
\eqref{BulkPerturbationEq:scalar3} using
\eqref{BulkPerturbationEq:scalar2} and
\eqref{BulkPerturbationEq:scalar4}, we obtain
\Eq{
\frac{1}{r^{n+1}}D_a(r^{n+1}\Sigma^a)
-\frac{k}{r}\Sigma
+\frac{n-1}{n}\frac{k^2-nK}{kr}\tau_T
+\frac{k}{2r}({\bar T}^{ab}F_{ab}-{\bar P}F^a_a)=0.
\label{BulkPerturbationEq:EM1}
}
This is just the equation $\delta(\bar\nabla_M {\bar
T}^M_i)=0$. Similarly, applying the same procedure to the divergence
of \eqref{BulkPerturbationEq:scalar1}, we obtain the equation
$\delta(\bar\nabla_M{\bar T}^M_a)=0$ which is expressed as
\Eq{
\frac{1}{r^n}D_b\left[r^n(\Sigma^b_a-{\bar T}^c_a F^b_c)\right]
+\frac{k}{r}\Sigma_a-n\frac{D_a r}{r}\Sigma 
 +{\bar T}_a^b D_bF-{\bar P}D_aF
+\frac{1}{2}\left({\bar T}^b_a D_bF^c_c
-{\bar T}^{bc}D_a F_{bc}\right)=0.
\label{BulkPerturbationEq:EM2}
}
%
Thus, naively speaking, only $m(m-1)/2$ components of
\eqref{BulkPerturbationEq:scalar1} are independent under
\eqref{BulkPerturbationEq:scalar2} and 
\eqref{BulkPerturbationEq:scalar4}, provided that the bulk 
energy-momentum conservation laws \eqref{BulkPerturbationEq:EM1}
and \eqref{BulkPerturbationEq:EM2} are satisfied. However, it is in
general difficult to extract such component explicitly.


\section{Junction Condition}
\label{sec:JunctionCondition}

In the brane-world model the bulk spacetime $\M$ has one or two
boundaries, and we live in a boundary $\Sigma$. Hence the intrinsic
geometry of $\Sigma$ is determined by the continuity of the bulk
metric ${\bar g}_{MN}$ and is described by the induced metric
$g_{\mu\nu}$. The intrinsic metric $g_{\mu\nu}$ determined in this
way, however, is dependent on the location of the boundary $\Sigma$ in
the bulk spacetime even if the geometry of the bulk spacetime is
given. Furthermore, in the spacetime with boundaries, the bulk
geometry is not uniquely determined by an initial condition unless
some appropriate boundary condition is imposed at $\Sigma$. Thus, in
order for the brane-world model to be well formulated, we must give
some additional prescription to determine the motion of branes and the
boundary condition at the branes for the bulk geometry.

In the brane-world models proposed so far, this prescription is
obtained by assuming that the bulk spacetime with boundaries is 
obtained from a spacetime $\tilde \M$ with $\ZR_2$ symmetry by
identifying points connected by the corresponding $\ZR_2$
transformation. The boundaries correspond to fixed points of the
transformation in the original covering spacetime $\tilde \M$. This
implies that the hypersurface in $\tilde \M$ corresponding to a
boundary $\Sigma$ is in general a singular surface in the sense that
the extrinsic curvatures $K_{\mu\nu}$ of $\Sigma$ on its two sides
have the same absolute value but their signs are different. Such a 
singular spacetime is obtained when the surface has an intrinsic
energy-momentum with finite surface density $T_{\mu\nu}$.

As is shown by Israel\cite{Israel.W1966}, this energy-momentum 
surface density is related to the difference of the extrinsic 
curvature on the two sides of the singular surface $\Sigma$. If we 
define $K_{\mu\nu}$ in terms of the unit normal $n_M$ to $\Sigma$ as
\Eq{
K_{\mu\nu}=-\bar \nabla_\mu n_\nu,
}
and denote its value on the side in the direction of $n^M$ as 
$K_+{}_{\mu\nu}$ and that on the other side as $K_-{}_{\mu\nu}$, 
this relation is written as
\Eq{
K_+{}^\mu_\nu-K_-{}^\mu_\nu=\kappa^2\left(T^\mu_\nu-\frac{1}{n}T 
\delta^\mu_\nu\right),
}
where the dimension of $\Sigma$ is $n+1$. In the brane-world model, 
if we choose the normal vector so that it is directed toward the 
inside of the bulk spacetime, 
$K_+{}^\mu_\nu=-K_-{}^\mu_\nu=K^\mu_\nu$. Hence the junction 
condition can be rewritten as
\Eq{
\kappa^2 T^\mu_\nu=2(K^\mu_\nu -K\delta^\mu_\nu).
\label{JC:general}
}
Thus, when the intrinsic dynamics of matter in the brane is given, 
the motion of brane is constrained by this junction condition. 

In this section we express the perturbation of the above junction
condition in terms of gauge-invariant variables. We consider only 
the case in which the unperturbed geometry of the brane is spatially
homogeneous and isotropic. This implies the case $m=2$ for the bulk
spacetime, i.e., $\M=\N^2\times \K^n$ locally, and the brane is
represented by a manifold
\Eq{
\Sigma=\RF\times \K^n\ni(\tau,x^i)=(x^\mu),
}
where $\K^n$ corresponds to the maximally symmetric space in the 
unperturbed background.


\subsection{Constraints}

The junction condition \eqref{JC:general} together with the
Hamiltonian constraint and the momentum constraint for the bulk
spacetime gives relations between quantities intrinsic to the brane
and the bulk energy-momentum density.  First, from the momentum
constraint
\Eq{
\nabla_\nu(K^\nu_\mu-K\delta^\nu_\mu)
=-\kappa^2 {\bar T}_{\mu\orth},
\label{MomentumConstraint:general}}
where $\nabla$ is the covariant derivative with respect to the 
induced metric $g_{\mu\nu}$ on $\Sigma$, and $\orth$ denotes the 
component along $n$, we obtain
\Eq{
\nabla_\nu T^\nu_\mu=-2{\bar T}_{\mu \orth}.
\label{EMconservation:general}}
Thus when the bulk spacetime is vacuum, the intrinsic 
energy-momentum tensor is conserved. 

Secondly, from the Hamiltonian constraint
\Eq{
K^2 - K^\mu_\nu K^\nu_\mu - R
=2\kappa^2 {\bar T}_{\orth\orth}-2\Lambda,
\label{HamiltonianConstraint:general}}
where $R$ is the Ricci scalar of $\Sigma$, we obtain
\Eq{
-R-
\frac{\kappa^4}{4}\left(T^\mu_\nu T^\nu_\mu-\frac{1}{n}T^2\right)
=2\kappa^2 {\bar T}_{\orth\orth}-2\Lambda.
\label{ScalarCurvatureConstraint:general}}
This implies that the expansion law of the brane universe is 
different from the one without the extra-dimension for which the 
relation
\Eq{
(n-1)R=-2\kappa^2 T
}
holds if the cosmological constant is included in $T_{\mu\nu}$.


\subsection{Unperturbed brane motion}

In the unperturbed background the brane motion is described by the
dependence of the $y^a$ coordinates on the proper time $\tau$ of
$\Sigma$, i.e., the set of functions $y^a(\tau)$. We define the unit
time-like vector $u^a$ by $u^a=\dot y^a$. Here and from now on the
over dot denotes the differentiation with respect to the proper time
$\tau$. The unit normal to $\Sigma$ in the unperturbed background is
uniquely determined by $u$ as
\Eq{
 n_a=-\epsilon_{ab}u^b,\ u_a=-\epsilon_{ab}n^b.
}

The extrinsic curvature is calculated as
\Eq{
K_{\tau\tau}= n_b u^aD_a u^b,\  
K_{\tau i}=0, \
K^i_j=-\frac{D_\orth r}{r}\delta^i_j,
}
and the unperturbed energy-momentum tensor of the brane is written 
as
\Eq{
T_{\tau\tau}=\rho, \ 
T_{\tau i}=0,\ 
T^i_j=p \delta^i_j.
}
Hence the junction condition is expressed as
\Eqr{
&& \frac{D_\orth r}{r}=-\frac{\kappa^2}{2n}\rho,
\label{JC:BG1}\\
&& (n-1)\frac{D_\orth r}{r}-K^\tau_\tau=\frac{\kappa^2}{2}p.
\label{JC:BG2}
}
The first of these equations implies that the energy density of our
universe is determined by the brane motion. If the equation of state
of the cosmic matter is given, these equations determine the brane
motion because $K^\tau_\tau$ represents the acceleration of the brane. 
Further, by differentiating the first equation by $\tau$ and
eliminating $K^\tau_\tau$, we obtain
\Eq{
\dot\rho + n(\rho+p)\frac{\dot a}{a}
=2u^a{\bar T}_{a \orth}. 
\label{EC:BG}}
This equation coincides with \eqref{EMconservation:general} 
obtained from the momentum constraint. 
Here $a$ denotes the value of $r$ at the brane and represents the 
cosmic scale factor of the Robertson-Walker universe on the brane 
whose metric is written as 
\Eq{
ds^2 = g_{\mu \nu}dx^\mu dx^\nu = - d\tau^2 + a^2(\tau) d\sigma_n^2 . 
}


\subsection{Perturbation of the Junction Condition}

The extrinsic curvature of the brane depends on the configuration 
of the brane as well as on the bulk geometry. If we denote the 
deviation of the brane configuration from the background one as
\Eq{
\delta z^M=Z^M(\tau,x) = Z^M_\para + Z_\orth n^M,
}
where $Z^M_\para$ is the component of $Z^M$ parallel to the brane, 
the perturbation of the extrinsic curvature is in general expressed 
as
\Eqr{
&\delta K_{\mu\nu}=
& (\Lie_{Z_\para}K)_{\mu\nu}+\nabla_\mu\nabla_\nu Z_\orth 
 + ({\bar R}_{\orth\mu \orth\nu}-K_\mu^\lambda 
K_{\lambda\nu})Z_\orth \nonumber\\
&& + n_a\delta {\bar \Gamma}^a_{\mu\nu} + \frac{1}{2}h_{ab}n^an^b 
K_{\mu\nu}.
\label{Perturbation:ExtrinsicCurvature:general}}

The perturbation of the intrinsic metric of the brane also depends 
both on the perturbation of the bulk metric and on the brane 
configuration. To be explicit, these relations are expressed as
\Eqr{
&& \delta g_{\tau\tau}=h_{ab}u^au^b-2\dot Z^\tau+2K^\tau_\tau 
Z_\orth,\\&& 
\delta g_{\tau i}=h_{ai}u^a-\hat D_iZ^\tau+a^2\Tdot{(Z_i/a^2)},\\
&& 
\delta g_{ij}=h_{ij}+\hat D_i Z_j+\hat D_j Z_i
+2a^2\gamma_{ij}\frac{D_ar}{r}Z^a.
}

To proceed further, we must treat the tensor, the vector and the 
scalar perturbation separately.


\subsubsection{Tensor perturbation}

For the tensor perturbation the perturbation of the intrinsic metric 
of the brane is expanded in terms of the tensor harmonics as
\Eq{
\delta g_{\tau\tau}=0,\ 
\delta g_{\tau i}=0, \
\delta g_{ij}=2a^2 h_T \THB_{ij}.
}
Since $Z^M=0$ for the tensor perturbation, $h_T$ is simply related 
to the bulk perturbation as $h_T=H_T$. 

The perturbation of the energy-momentum tensor intrinsic to the 
brane is also expressed by a single expansion coefficient 
representing the anisotropic stress perturbation of the brane as
\Eq{
\delta T^\tau_\tau=0,\ 
\delta T^\tau_i=0,\
\delta T^i_j = \pi_T \THB^i_j.
}
On the other hand, the harmonic expansion of 
\eqref{Perturbation:ExtrinsicCurvature:general} yields
\Eq{
\delta K^\tau_\tau=0,\ 
\delta K^\tau_i=0,\ 
\delta K^i_j=-D_\orth H_T \THB^i_j.
}
Hence the junction condition \eqref{JC:general} 
reduces to the single equation
\Eq{
D_\orth H_T=-\frac{\kappa^2}{2}\pi_T.
\label{JC:tensor}}

In general, the anisotropic stress perturbation is not an independent
dynamical variable and is expressed by other dynamical variables when
the model is specified. In particular, in the linear perturbation
framework, it is natural to assume that $\pi_T=0$ for the tensor
perturbation. In this case \eqref{JC:tensor} gives a Neumann-type
boundary condition for the wave equation of $H_T$ obtained in Section
\ref{sec:BulkPerturbation:tensor}. Thus we obtain a well-posed system
describing the evolution of perturbations.


\subsubsection{Vector perturbation}

For the vector perturbation the perturbation of the brane 
configuration is expressed in the harmonic expansion as
\Eq{
 Z^\tau=0, \ Z_\orth=0, \ Z_i=aZ \VHB_i.
}
On the other hand the intrinsic metric perturbation is expressed as
\Eq{
\delta g_{\tau\tau}=0,\ 
\delta g_{\tau i}=-a\beta \VHB_i, \
\delta g_{ij}=2a^2 h_T \VHB_{ij}.
}
Hence we obtain the relations
\Eqr{
&& \beta=-f_\para -a\Tdot{\Frac(Z/a)},\\
&& h_T=H_T-\frac{k}{a}Z.
}

If we construct the standard gauge-invariant variables for the 
intrinsic perturbation from these metric perturbation variables and 
the matter perturbation variables defined by
\Eq{
\delta T^\tau_\tau=0,\ 
\delta T^\tau_i=a(\rho+p)(v-\beta)\VHB_i,\
\delta T^i_j = \pi_T \VHB^i_j,
}
we obtain
\Eqr{
&& \sigma_g=\frac{a}{k}\dot h_T-\beta= F_\para,\\
&& V=v-\beta=v-\frac{a}{k}\dot h_T+F_\para.
}
Note that $Z$ disappears in these expressions because it corresponds
to an intrinsic diffeomorphism of the brane. On the other hand, 
in the present case the perturbation of the extrinsic curvature is 
expressed as
\Eq{
\delta K^\tau_\tau=0,\ 
\delta K^\tau_i=\frac{a^2}{2}\epsilon^{ab}D_a\Frac(F_b/r) \VHB_i,\ 
\delta K^i_j=-\frac{k}{a}F_\orth \VHB^i_j.
}
Inserting these equations into \eqref{JC:general}, we 
obtain the following two equations:
\Eqr{
&& \kappa^2(\rho+p)V=r\epsilon^{ab}D_a\Frac(F_b/r),
\label{JC:vector1}\\
&& \kappa^2 \pi_T= -2\frac{k}{a}F_\orth.
\label{JC:vector2}}
The first of these gives the expression for the intrinsic perturbation
variable in terms of the bulk perturbation variable. 
The second can be regarded as the boundary condition on the bulk
perturbation equations in Section
\ref{sec:BulkPerturbation:vector}. It will be shown later that it 
gives a Dirichlet-type boundary condition when the bulk spacetime is 
vacuum.


\subsubsection{Scalar perturbation}
\label{sec:JC:scalar}

For the scalar perturbation for which 
\Eq{
Z^\tau=Z^\tau \SHB,\ 
Z_\orth=Z_\orth \SHB,\ 
Z_i=a Z \SHB_i,
}
the harmonic expansion coefficients for the intrinsic metric 
perturbation defined by
\Eq{
\delta g_{\tau\tau}=-2\alpha\SHB,\ 
\delta g_{\tau i}=-a\beta \SHB_i, \
\delta g_{ij}=2a^2 ( h_L\SHB\gamma_{ij}+h_T \SHB_{ij}),
}
are related to those for the bulk metric perturbation as
\Eqr{
&& \alpha=-\frac{1}{2}F_{\para\para}+\dot Y^\tau
-K^\tau_\tau Y_\orth,\\
&& \beta=-\frac{k}{a}Y^\tau+\frac{a}{k}\dot h_T,\\
&& h_L=H_L+\frac{k}{na}Z+\frac{\dot a}{a}Z^\tau
+\frac{D_\orth r}{r}Z_\orth,\\
&& h_T=H_T-\frac{k}{a}Z,
}
where 
\Eq{
Y^\tau=Z^\tau-X^\tau,\ 
Y_\orth=Z_\orth-X_\orth.
}
Hence the intrinsic gauge-invariant variables constructed from these 
are related to the bulk gauge-invariant variables as
\Eqr{
&& \Phi=h_L+\frac{1}{n}h_T-\frac{\dot a}{k}\sigma_g
=F+\frac{D_\orth r}{r}Y_\orth, 
\label{PhiByF}\\
&& \Psi=\alpha-\frac{1}{k}\Tdot{(a\sigma_g)}
= -\frac{1}{2}F_{\para\para}-K^\tau_\tau Y_\orth,
\label{PsiByF}}
where
\Eq{
\sigma_g=\frac{a}{k}\dot h_T-\beta.
}

In addition to these, we can construct gauge-invariant variables 
from the harmonic expansion of the intrinsic matter perturbation
\Eq{
\delta T^\tau_\tau=-\delta\rho \SHB,\ 
\delta T^\tau_i=a(\rho+p)(v-\beta)\SHB_i,\
\delta T^i_j = \delta p \SHB\delta^i_j+ \pi_T \SHB^i_j,
}
as
\Eqr{
&& V=v-\frac{a}{k}\dot h_T,\\
&& \rho\Delta=\delta \rho -\frac{a}{k}\dot\rho (v-\beta),\\
&& \Gamma=\delta p -c_s^2\delta\rho.
}
Among these the last one represents the amplitude of entropy 
perturbation of the matter.

The perturbation of the extrinsic curvature are now expressed in 
terms of the gauge-invariant variables as
\Eqr{
&\delta K^\tau_\tau= 
&\left[-\frac{1}{2}\dot F_{\orth\para}+\frac{1}{2}n_aD_b 
F^{ab}-\frac{1}{2}D_\orth F^a_a +\frac{1}{2}K^\tau_\tau 
F_{\para\para} \right.\nonumber\\
&&\left.+\dot K^\tau_\tau Y^\tau-\ddot Y_\orth
+\left(\frac{1}{2}\,{}^2\!R+K_{\tau\tau}^2\right)Y_\orth
\right]\SHB,\\
&\delta K^\tau_i=
& k\left[\frac{1}{2}F_{\orth\para}
-\left(K^\tau_\tau+\frac{D_\orth r}{r}\right)Y^\tau
+a\Tdot{\Frac(Y_\orth/a)}\right]\SHB_i,\\
&\delta K^i_j=
& \left[-D_\orth F-\frac{\dot a}{a}F_{\orth\para}
+\frac{D_\orth r}{2r}F_{\orth\orth}
-\Tdot{\Frac(D_\orth r/r)}Y^\tau \right.\nonumber\\
&& \left. -\frac{\dot a}{a}\dot Y_\orth
-\left(\frac{k^2}{na^2}+\frac{n^an^bD_aD_b r}{r}
-\Frac(D_\orth r/r)^2\right)Y_\orth\right]\SHB\delta^i_j \nonumber\\
&& +\frac{k^2}{a^2}Y_\orth \SHB^i_j.
}
Hence the junction condition \eqref{JC:general} 
yields the following three relations among the gauge-invariant 
variables for the bulk and the brane:
\Eqr{
&& D_\orth F+\frac{\dot a}{a}F_{\orth\para}
-\frac{D_\orth r}{2r}F_{\orth\orth} \nonumber\\
&& \quad +\frac{\dot a}{a}\dot Y_\orth
+\left(\frac{k^2}{na^2}+\frac{n^an^bD_aD_b r}{r}
-\Frac(D_\orth r/r)^2\right)Y_\orth \nonumber\\
&&\quad =-\frac{\kappa^2}{2n}\left(\rho\Delta+\frac{a}{k}\dot \rho 
V\right),
\label{JC:scalar1}\\
&& -\frac{1}{2}\dot F_{\orth\para}+\frac{1}{2}n_aD_bF^{ab}
-\frac{1}{2}D_\orth F^a_a + \frac{1}{2}K^\tau_\tau F_{\para\para} 
\nonumber\\
&& \quad -\ddot Y_\orth
+\left(\frac{1}{2}\,{}^2\!R+K_{\tau\tau}^2\right)Y_\orth \nonumber\\
&&\quad 
=-\frac{\kappa^2}{2}\left[\Gamma+\left(\frac{n-1}{n}+c_s^2\right)\left
(\rho\Delta+\frac{a}{k}\dot \rho V \right)\right],
\label{JC:scalar2}\\
&& \frac{1}{2}F_{\orth\para}+a\Tdot{\Frac(Y_\orth/a)}
=\frac{\kappa^2}{2}\frac{a}{k}(\rho+p)V,
\label{JC:scalar3}\\
&& 2\frac{k^2}{a^2}Y_\orth=\kappa^2\pi_T.
\label{JC:scalar4}}

These conditions have some features that are not shared by the 
vector and tensor perturbation. First, although the variables 
$Z^\tau$ and $Z$ disappear as in the other types of perturbation, 
$Y_\orth=Z_\orth -X_\orth$ remains in the final expressions. This 
is because $Y_\orth$ defines the gauge-invariant amplitude of 
the perturbation of the brane motion, unlike $Z^\tau$ and $Z$, 
which correspond to intrinsic diffeomorphism of the brane. 
Secondly, from the last equation one finds that a condition on 
the anisotropic stress perturbation does not give any boundary condition 
on the bulk perturbation. 
Instead, it constrains the perturbation which cannot 
be simply attributed either to the intrinsic structure of the brane 
or to the bulk.

Then where does the boundary condition comes from? We can find an answer 
to this question by closely inspecting the structure of the above 
equations. First, note that the gauge-invariants representing the 
perturbation of the intrinsic geometry of the brane are determined 
by the bulk variables through \eqref{PhiByF} and \eqref{PsiByF}. 
Meanwhile, \eqref{JC:scalar1} and \eqref{JC:scalar3} yield the 
expressions of the gauge-invariants $\Delta$ and $V$ for the 
intrinsic matter in terms of the bulk variables. Inserting these 
expressions into \eqref{JC:scalar2}, we obtain an expression for the 
amplitude of the entropy perturbation $\Gamma$ in terms of the bulk 
variables. Like $\pi_T$, $\Gamma$ is not a dynamical variable and 
should be expressed in terms of $\Delta$, $V$ and other intrinsic 
dynamical perturbation variables whose dynamics is determined when a 
model of the intrinsic matter is given. Hence we should regard 
\eqref{JC:scalar2} or an equation derived from it by eliminating the 
independent dynamical variable as the boundary condition on the bulk
perturbation. This means that the boundary condition is dependent on
the type of the intrinsic matter perturbation, e.g., adiabatic or
isocurvature. In the next section we will show that this boundary
condition becomes non-local with respect to the time coordinate of the
brane.


\section{Master Variable}

As was shown in \ref{sec:BulkPerturbationEq}, the metric perturbation
in the bulk spacetime for the tensor perturbation is described by the
single gauge-invariant variable $H_T$, and it obeys a simple wave
equation. Further the junction condition gives a simple boundary
condition on it. In contrast, for the vector and the scalar
perturbations, the bulk perturbation is described by multi-component
variables and their equations have structures too complicated to be
solved. Fortunately, in the case in which the unperturbed background
of the bulk spacetime is vacuum ( and the 2-dimensional orbit space 
${\cal N}^2$ is maximally symmetric for the
scalar perturbation), we can find a single master variable for the
bulk perturbation and reduce the perturbation equation to a single
wave equation. In this section we analyze the structure of the
junction condition in terms of that master variable.


\subsection{Vacuum Background}

We consider the case in which $m=2$ in the notation of Section
\ref{sec:BulkBackground} and ${\bar T}_{MN}=0$.  Hence the bulk
spacetime is $(n+2)$-dimensional and has the isometry group
corresponding to the $n$-dimensional maximally symmetric space in the
unperturbed background. In this case, from the generalized Birkhoff
theorem, the geometry of the background spacetime is given by either
of the following two families of solutions:
\begin{itemize}
\item[] 1) Pure product type $(Dr=0)$:
\Eqr{
& \Lambda>0 : & dS^2(\sqrt{n/2\Lambda})\times S^n(\sqrt{n(n-1)/2\Lambda}),\\
& \Lambda<0 : & AdS^2(\sqrt{n/2|\Lambda}|)\times H^n(\sqrt{n(n-1)/2|\Lambda|}),\\
& \Lambda=0: & E^{n+1,1}.
}
\item[] 2) Schwarzschild type$(Dr\not=0)$:
\Eqr{
&& d\bar{s}^2=-U(r)dt^2+\frac{dr^2}{U(r)}+r^2d\sigma_n^2,
\label{BGsol:S}\\
&& U(r)=K-\frac{2M}{r^{n-1}}-\lambda r^2;\\
&& \lambda=\frac{2\Lambda}{n(n+1)}.
}
\end{itemize}
The derivation of the solutions of the first family and their 
physical meaning were given by Nariai~\cite{Nariai.H1950,Nariai.H1951}. 
For the second family the following simple formulas hold:
\Eqr{
&&{}^2\!R=2\lambda+\frac{2n(n-1)M}{r^{n+1}},\\ 
&& \frac{\square r}{r}=-2\lambda+\frac{2(n-1)M}{r^{n+1}},\\
&& \frac{K-(Dr)^2}{r^2}=\lambda+\frac{2M}{r^{n+1}}
\label{Dr:BH:S}.
}
In particular, when the mass parameter $M$ vanishes, the quantities on
the left-hand side of these equations become constant, and the
spacetime coincides with $dS^{n+2}$, $AdS^{n+2}$ and $E^{n+1,1}$ for
$\Lambda>0, <0$, and $=0$, respectively.

The background configuration of the brane in the Schwarzschild-type
background geometry is determined by solving \eqref{JC:BG1} and
\eqref{JC:BG2} with \eqref{Dr:BH:S}. In particular, from \eqref{EC:BG}, the 
same energy equation as in the no-extradimension case holds for the
energy density of the brane,
\Eq{
\dot\rho =-n(\rho+p)\frac{\dot a}{a}.
\label{EC:BG:vacuum}}
In contrast, from \eqref{JC:BG1} and \eqref{Dr:BH:S}, the
decomposition $(Dr)^2=-\dot a^2+(D_\orth r)^2$ yields
\Eq{
\Frac(\dot a/a)^2=\left(\frac{\kappa^2}{2n}\rho\right)^2
-\frac{K}{a^2}+\lambda+\frac{2M}{a^{n+1}},
}
which is different from the standard expansion equation even in the
case $M=0$ in the point that $\rho$ is replaced by $\rho^2$.  These
equations form a closed system and determine $\rho$ and $a$ as
functions of the intrinsic proper time $\tau$. When these functions
are given, the embedding of the brane, $(t(\tau), r(\tau))$ is
determined by $r(\tau)=a(\tau)$ and a solution of the equation
\Eq{
\Frac(dt/d\tau)^2=\frac{\dot a^2+U(a)}{U(a)^2}.
}

In contrast to the Schwarzschild case, the background brane
configuration becomes quite special for the pure product type
background spacetime. In fact, since $r$=const in this case, it
follows from \eqref{JC:BG1} and \eqref{JC:BG2} that $\rho$ should
vanish and $K^\tau_\tau$ is proportional to $p$. Since it is natural
to assume $p=0$ for $\rho=0$, the latter condition implies that the
background brane motion is represented by a geodesic in the
2-dimensional constant curvature space $\N$.


\subsection{Expression in terms of a master variable}


\subsubsection{Tensor perturbation}

For the tensor perturbation the system is already described by a
single variable. For completeness we recapitulate the equations for
the tensor perturbation in the vacuum case. We need no further
symmetry assumption on the unperturbed bulk geometry.

The perturbation equation for the bulk is given by the
homogeneous wave equation
\Eq{
-\square H_T-\frac{n}{r}Dr\cdot DH_T+ \frac{k^2+2K}{r^2}H_T
=0.
}
The junction condition gives the boundary condition
\Eq{
D_\orth H_T=-\frac{\kappa^2}{2}\pi_T.
}
%


\subsubsection{Vector perturbation}

For the vector perturbation on the vacuum bulk spacetime $\tau_T$
vanishes. Hence for $k^2>(n-1)K$, taking account of the fact that the
orbit space $\N$ is two-dimensional,
\eqref{BulkPerturbationEq:vector2} implies that $F_a$ is 
written in terms of a function $\Omega$ as
\Eq{
F^a=\frac{1}{r^{n-1}}\epsilon^{ab}D_b\Omega.
}
Hence the perturbation equation \eqref{BulkPerturbationEq:vector1} for
$F_a$ is expressed in terms of $\Omega$ as
\Eq{
D^a\left[r^{n+2}D_b\Frac(D^b\Omega/r^n) -\left\{k^2-(n-1)K\right\}\Omega
\right]=0.
}
The bulk perturbation equation is thus reduced to the single equation
for the master variable $\Omega$ given by
\Eq{
\square \Omega - \frac{n}{r}Dr\cdot D\Omega 
-\frac{k^2-(n-1)K}{r^2}\Omega=\frac{C}{r^2},
}
where $C$ is an integration constant, which can be set to zero by 
redefinition of $\Omega$.

On the other hand, for the mode $k^2=(n - 1)K>0$, the gauge-invariant 
$F^{(1)}_{ab}$ has a single independent component and is expressed as
\Eq{
F^{(1)}_{ab}=\epsilon_{ab}\Omega^{(1)}.
}
In terms of $\Omega^{(1)}$ 
\eqref{BulkPerturbationEq:vector3} is expressed as
\Eq{
\epsilon_{ab}D^b\left(r^{n+1}\Omega^{(1)}\right)=0.
}
This equation is easily solved to yield
\Eq{
F^{(1)}_{ab}=\epsilon_{ab}\frac{C}{r^{n+1}},
}
where $C$ is an integration constant.

For $k^2>(n-1)K$, the junction conditions \eqref{JC:vector1} and 
\eqref{JC:vector2} are expressed in terms of $\Omega$ as
\Eqr{
&& \kappa^2\pi_T= -2\frac{k}{a^n}\dot \Omega,\\
&& \kappa^2(\rho+p)V=\frac{1}{a^{n+1}}[k^2-(n-1)K]\Omega,\\&& 
\sigma_g=\frac{1}{a^{n-1}}D_\orth \Omega.
}
The first equation gives a Dirichlet-type boundary condition on
$\Omega$. The other two equations give expressions for the intrinsic 
gauge-invariant variables $V$ and $\sigma_g$ in terms of $\Omega$. 
Thus the initial value problem is well-posed for this system.

The situation for the exceptional mode $k^2=(n-1)K>0$ is slightly
different. For this mode we do not have the equation for
$\pi_T$. However, this does not cause trouble because $F^{(1)}_{ab}$
is explicitly given. The junction condition determines the only
non-trivial gauge-invariant intrinsic to the brane, $V$, as
\Eq{
\kappa^2a^{n+1}(\rho+p)V=C.
}

Here note that the momentum constraint 
\eqref{EMconservation:general} reduces to the conservation 
of $T_{\mu\nu}$ in the present case and its perturbation gives
\Eq{
\frac{1}{a^{n+1}}\Tdot{\left[a^{n+1}(\rho+p)V\right]}
=\frac{k^2-(n-1)K}{2ak}\pi_T.
}
It is easily checked that this equation is consistent with the above
junction conditions. Thus the evolution of $V$ is intrinsically
determined and coincides with the no-extra-dimension case. In contrast,
the evolution of $\sigma_g$ is determined only by solving the master
equation, in contrast to the no-extra-dimension case in which $\sigma_g$ is
related to $V$ as\cite{Kodama.H&Sasaki1984}
\Eq{
2{\kappa'}^2 a^2(\rho+p)V=-[k^2-(n-1)K]\sigma_g , 
}
where ${\kappa'}^2$ denotes the gravitational constant on the brane. 


\subsubsection{Scalar perturbation}

As shown in Section \ref{sec:BulkPerturbation:scalar}, 
for the scalar perturbation on the vacuum background, 
\eqref{BulkPerturbationEq:scalar3} is automatically 
satisfied if the other three hold. Among the latter, 
\eqref{BulkPerturbationEq:scalar2} and 
\eqref{BulkPerturbationEq:scalar4} are written as
\Eqr{
&& F^a_a=-2(n-2)F,
\label{BulkPEq:vacuum1}\\
&& D_b(r^{n-2} F^b_a)=2D_a(r^{n-2}F).
\label{BulkPEq:vacuum2}
}
Here note that for the exceptional modes $k^2=0$ and $k^2=nK>0$ we 
do not have one or both of them. However, we can still assume that 
these equations hold by regarding missing equations as gauge 
conditions to fix the residual gauge freedom.

As was shown by Mukohyama, in the case that the two-dimensional 
orbit space $\N$ is a constant curvature space, the 
general solutions to these equations are written in terms 
of a master variable $\Omega$ as
\Eqr{
&& \tilde F=r^{n-2}F=\frac{1}{2n}(\square +2\lambda)\Omega,
\label{MV:maxsymm1}\\
&& \tilde F_{ab}=r^{n-2}F_{ab}=D_aD_b\Omega
-\left(\frac{n-1}{n}\square+\frac{n-2}{n}\lambda\right)\Omega g_{ab}.
\label{MV:maxsymm2}
}
(see Appendix \ref{Appendix:C} for a simpler proof.)  

On the other hand, for the background geometry 
\eqref{BGsol:S} with $M=0$, 
\eqref{BulkPerturbationEq:scalar1} is reduced to the 
following equation:
\Eqr{
&& -\square F_{ab}-\frac{n}{r}Dr\cdot DF_{ab}
+\left(\frac{k^2}{r^2}-2\lambda\right)F_{ab}\nonumber\\
&&\quad +\frac{D^cr}{r}\left[
2(D_a F_{cb}+D_b F_{ca})
+(n-2)\left(\frac{D_ar}{r}F_{cb}+\frac{D_br}{r}F_{ca}\right)\right]
\nonumber\\
&&\quad =4\left[\frac{D_b r}{r}D_aF+\frac{D_ar}{r}D_bF
+(n-2)\frac{D_arD_br}{r^2}F\right].
}
In terms of the master variable $\Omega$, this equation is written as
\Eq{
(D_aD_b+\lambda g_{ab})E(\Omega)=0,
\label{PDE}}
where
\Eq{
E(\Omega)\equiv r^2\left[\square \Omega -\frac{n}{r}Dr\cdot D\Omega
-\left\{\frac{k^2-nK}{r^2}+(n-2)\lambda\right\}\Omega\right].
}

As is shown in Appendix \ref{Appendix:D}, the general solution of
\eqref{PDE} is written as
\Eq{
E(\Omega)=C_0 g_0(t,r)+C_1 g_1(t,r) + C_2 r,
}
where $C_0, C_1$ and $C_2$ are arbitrary constants. On the other hand,
it is easy to see that the freedom in the definition of $\Omega$ is
expressed in terms of a solution to $(D_aD_b+\lambda g_{ab})\omega=0$
as $\Omega \tend \Omega + \omega$. Since $\omega$ is again written as
$\omega=C'_0 g_0(t,r)+C'_1 g_1(t,r) + C'_2 r$ with arbitrary constants
$C'_0\sim C'_2$, the value of $E(\Omega)$ changes by the redefinition
as
\Eq{
E(\omega)=\left\{\begin{array}{ll}
  -(k^2-nK)(C'_0 g_0+C'_1 g_1)-k^2C'_2 r; & K\not=0\\
  -k^2(C'_0 g_0+C'_1 g_1)-(k^2C'_2-2n\lambda C'_0)r; &K=0
  \end{array}\right.
}

From this we immediately see that $C_0\sim C_2$ can be set to zero by
an appropriate redefinition of $\Omega$ for $k^2(k^2-nK)\not=0$. On
the other hand, only $C_0$ and $C_1$ can be put to zero for $k^2=0$
and $K\not=0$, while only $C_2$ can be put to zero for $k^2=nK>0$. In
these cases, however, there still remains a residual gauge freedom in
$F$ and $F_{ab}$. As is shown in Appendix \ref{Appendix:E}, any
solution $\Omega$ to the homogeneous equation $E(\Omega)=0$ can be set 
to zero by this residual gauge transformation, while the constants above 
that cannot be removed by the redefinition are just the
gauge invariants for the exceptional modes. Thus the gauge equivalent
classes of the solutions to the perturbed solutions form a
one-dimensional space parametrized by $C_2$ for the mode $k^2=0$ and
$K>0$ and a two-dimensional space parametrized by $C_0$ and $C_1$ for
the mode $k^2=nK>0$.

From now on we consider only modes with $k^2(k^2-nK)\not=0$. From the
above argument, the master equation for these modes is always written
as
\Eq{
\square \Omega -\frac{n}{r}Dr\cdot D\Omega
-\left[\frac{k^2-nK}{r^2}+(n-2)\lambda\right]\Omega
=0.
}

In terms of the master variable, the junction conditions
\eqref{JC:scalar1}$\sim$\eqref{JC:scalar4} are written as
\Eqr{
&& rD_\orth\Frac(\Omega/r)=-\frac{\kappa^2}{k^2-nK}a^n\rho\Delta
-\frac{\kappa^2}{k^2}a^n\pi_T,
\label{JC:scalar:maxsymm1}\\
&& \Tdot{(D_\orth\Omega)}+K^\tau_\tau\dot\Omega
=\kappa^2\frac{a^{n-1}}{k}(\rho+p)V
-\frac{\kappa^2}{k^2}a^{n-1}\Tdot{(a\pi_T)},
\label{JC:scalar:maxsymm2}\\
&& \frac{1}{a}\Tdot{(aV)}=\frac{k}{a}\Psi
+\frac{k}{a}\frac{\Gamma+c_s^2\rho\Delta}{\rho+p}
-\frac{n-1}{n}\frac{k^2-nK}{ak}\frac{\pi_T}{\rho+p},
\label{JC:scalar:maxsymm3}\\
&& 2\frac{k^2}{a^2}Y_\orth=\kappa^2\pi_T.
\label{JC:scalar:maxsymm4}}
where
\Eqr{
&\Phi=
&\frac{1}{2a^{n-2}}\left[
-\frac{\dot a}{a}\dot\Omega+\frac{D_\orth r}{r}D_\orth\Omega
+\left(\frac{k^2-nK}{na^2}+\lambda\right)\Omega\right] \nonumber\\
&& +\frac{D_\orth r}{r}Y_\orth,\\
&\Psi=
&-\frac{1}{2a^{n-2}}\left[
\ddot \Omega +\left(K^\tau_\tau+(n-1)\frac{D_\orth r}{r}\right)D_\orth \Omega
-(n-1)\frac{\dot a}{a}\dot\Omega \right. \nonumber\\
&& \left. +\left(\frac{n-1}{n}\frac{k^2-nK}{a^2}
+(n-2)\lambda\right)\Omega\right]
-K^\tau_\tau Y_\orth.
}

Here note that \eqref{JC:scalar:maxsymm3} is identical to the space
component of the perturbation of the intrinsic conservation law of the
energy-momentum tensor, $\nabla_\mu T^\mu_\nu=0$. Further the
corresponding time component, which is written as
\begin{eqnarray}
 \frac{1}{a^n} \Tdot{(a^n\rho\Delta)} &=& {} \!
 -\frac{k}{a}(\rho+p)\left[1-n\frac{a^2}{k^2}\Tdot{\Frac(\dot a/a)}
\right]V -n(\rho+p)\left(\dot\Phi-\frac{\dot a}{a}\Psi\right) \nonumber\\
&& -(n-1)\frac{k^2-nK}{k^2}\frac{\dot a}{a}\pi_T,
\label{BraneECPT:scalar}
\end{eqnarray}
is obtained from the above junction conditions, as it should be.

As was discussed in Section \ref{sec:JC:scalar}, 
\eqref{JC:scalar:maxsymm1} and \eqref{JC:scalar:maxsymm2} 
are the equations determining the intrinsic gauge-invariants $\Delta$
and $V$. Hence the equation \eqref{JC:scalar:maxsymm3}, or the
equation for the intrinsic entropy perturbation $\Gamma$ should be
regarded as a boundary condition on the master variable. For
$\pi_T=0$, this expression is given by
\Eqr{
&& \Tddot{\left[rD_\orth\Frac(\Omega/r)\right]}
+ (2 + nc_s^2)\frac{\dot a}{a}\Tdot{\left[rD_\orth\Frac(\Omega/r)\right]}
\nonumber\\
&&\quad +\left\{-n(1+w)(2n-2+nw)\Frac(D_\orth r/r)^2
+c_s^2\frac{k^2-nK}{a^2}\right\}\left[rD_\orth \Frac(\Omega/r)\right]\nonumber\\
&&\quad - (n-1)(1+w)\frac{k^2}{a^2}\frac{D_\orth r}{r}\Omega 
=\kappa^2a^{n-2}\Gamma,
\label{BC:nonlocal}
}
where $w=p/\rho$. From this equation we immediately see 
that except for the special case in which $p=-\rho$, the 
junction condition yields a boundary condition that is non-local in time.

In contrast, for the case $p=-\rho$, the junction condition yields a
closed evolution equation for $rD_\orth(\Omega/r)$ or $\rho
\Delta$. 
To be precise, $\delta\rho$ and $\delta p$ becomes 
gauge invariant. Further, although $V$ is ill-defined, 
the combination $(\rho+p)V=(\rho+p)(v-\beta)=\delta T^\tau_i /(a\SHB_i)$ 
is well-defined and can have a non-vanishing value. 
If we take these facts into account, the boundary condition for 
$\rho+p=0$ is given by the equation obtained from (\ref{BC:nonlocal}) 
by the replacements $c_s^2=0$, $w=-1$ and $\Gamma=\delta p$.

Even in this case, the gauge invariants $\Phi$ and $\Psi$ representing
the intrinsic perturbations of the spatial curvature and the
gravitational potential of the brane are determined only by solving
the wave equation for $\Omega$ under given initial data and a boundary
condition. This is because we lack the relations that make the
equations for intrinsic quantities closed in the no-extra-dimension
case\cite{Kodama.H&Sasaki1984},
\Eqr{
&& {\kappa'}^2\rho\Delta=(n-1)a^{-2}(k^2-nK)\Phi,\\
&& (n-2)\Phi+\Psi=-{\kappa'}^2a^2 k^{-2}\pi_T.
}
Thus it may be difficult to find a natural initial condition for which
the evolution law for the intrinsic perturbation becomes similar to
the standard one.


\section{Discussion}

In the present paper we have developed a gauge-invariant formalism for
the perturbation of the brane-world model for which the background
configuration has a spatial symmetry corresponding to a maximally
symmetric space with a dimension $n$ lower than the dimension $n+m$ of
the bulk spacetime. The formalism consisted of two parts. The first
part gave a system of gauge-invariant equations for the perturbation
of the bulk spacetime geometry. With applications to wider situations
in mind, we derived the equations for generic values of $n$ and $m$
and for generic bulk matter. They give an extension of the formalism
developed for the $n=2$ and $m=2$ case by Gerlach and
Sengupta~\cite{Gerlach.U&Sengupta1979}.

The second part was concerned with a situation specific to 
the brane-world model in which $m=2$ and gave
gauge-invariant equations for the junction condition corresponding to
the $\ZR_2$ symmetry along a brane with codimension one.  As an
immediate consequence, we have shown that, when the stress perturbation
intrinsic to the brane is specified or expressed in terms of other
intrinsic quantities, the junction condition yields a boundary
condition at the brane(s) on the evolution equation for the bulk
perturbation.

In order to investigate the structure of the equations in more detail, we
have introduced a master variable $\Omega$ for the bulk perturbation
and reduced the bulk perturbation equations to a single wave equation
for $\Omega$ in the case in which the bulk spacetime is vacuum. This
reduction was already done by Mukohyama\cite{Mukohyama.S2000A} in the
case in which the background geometry of the bulk spacetime is
maximally symmetric. Since we were able to introduce the master
potential for the scalar perturbation only in the case in which the
two-dimensional orbit space has a constant curvature, the master
equation we obtained is the same as that derived by
Mukohyama. However, the master equation for the vector and tensor
perturbations is more general and holds also in the case in which the
background geometry is of the Schwarzschild black hole type. We have
also given a proof different from that given by Mukohyama for the
existence of the master potential for the scalar perturbation.

We have also investigated the structure of the junction condition in
terms of the master variable. In particular, we have shown that the
boundary condition on the master potential obtained from the junction
condition has a different structure depending on the type of
perturbation: for the tensor and vector perturbations, the
condition that the anisotropic stress perturbation vanishes yields a
Neumann-type and a Dirichlet-type boundary condition, respectively,
while the boundary condition for the scalar perturbation is given by a
condition on the intrinsic entropy perturbation and is non-local in
time in general.

Here, note that, although the master variable is used in an essential
way in the analysis of the scalar perturbation, the introduction of
the master variable is not the only way to make the problem tractable. 
For example, Fourier expansion of the original gauge-invariant
variables in terms of time may also be used to make the equations
simpler. If it works well, we can also treat the scalar perturbation
in the Schwarzschild black hole type background.

Although the main purpose of the present paper is to develop a
formalism, we briefly discuss here a possible consequence of the
formalism for the brane-world scenario. In the original Randall-Sundrum
model, in which the brane is realized as a flat subspace in a
5-dimensional Anti-de Sitter spacetime, the bulk graviton modes which
behave as massive particles inside the brane decouple from the
massless mode. In our formalism this phenomenon is understood in the
following way.

Since $n=3$, $K=0$ and $M=0$ in this case, in the units $\lambda=-1$,
the gravitational wave in the bulk spacetime is described by $H_T$
satisfying the wave equation
\Eq{
-\partial_t^2 H_T
=-\frac{1}{r}\partial_r\left(r\partial_r H_T\right)
+k^2 H_T.
}
Since the brane is static and located at $r=1$, the boundary 
condition is given by $\partial_r H_T=0$. Under the Fourier 
expansion with respect to the time $t$, the mode $H_T\propto 
e^{-i\omega t}$ is a solution to the equation
\Eq{
y^3\frac{d}{dy}\left(\frac{1}{y^3}\frac{dH_T}{dy}\right)
+\mu^2 H_T=0,
}
where $y=1/r$($1\le y<\infty$) and $\mu^2=\omega^2-k^2$. If we require
that the mode is normalizable in the generalized sense with respect to
the natural metric $dr r \propto dy/y^3$ which makes the right-hand
side of the above wave equation self-adjoint, the spectrum of $\mu^2$
consists of two parts. One is the point spectrum $\mu^2=0$ for which
$H_T$ is constant. The other is the continuous spectrum $\mu^2>0$ for
which $H_T$ is proportional to $y^2 Z_2(\mu y)$ where $Z_2$ is a
Bessel function of degree $2$. Thus the general solution is written as
\Eq{
H_T=\Re\left[C e^{-ikt}
+\int_0^\infty d\mu^2 y^2 \{ A(\mu)J_2(\mu y)+B(\mu)N_2(\mu y) \}
e^{-i\omega t} \right] .
}

The important point here is that the boundary condition is simply
written as a relation between $A$ and $B$. Hence the massless mode for
which $A=B=0$ decouples from massive modes. If we apply the same
argument to a dynamical case in which the brane is non-static and
represents an expanding universe, the situation changes significantly. 
In this case the boundary condition $D_\orth H_T=0$ is expressed as a
relation among $A$, $B$, and $C$. Hence all modes contain massive
components.

Of course, since the expansion rate of the present universe is small,
one might expect that there is a mode in which the amplitude of the
massive component is negligible. However, such a mode contains massive
components with large amplitudes in the early phase of the universe
due to rapid cosmic expansion. Hence, if the initial condition of the
universe is imposed in the early universe as in the argument of
quantum generation of perturbations, it is in general expected that
the present day universe contains a non-negligible amount of massive
gravitons. The situation is quite similar to the quantum particle
creation due to cosmic expansion. Whether this problem is a crucial
defect of the brane-world model, or it rather provides a new model of
dark matter is a very interesting problem.

\section*{Acknowledgments}

A.I. would like to thank Misao Sasaki for informative comments and
discussion.  A.I. was supported by Japan Society for the Promotion of
Science and in part by Handai Yukawa Shogakukai, and H.K was supported
by the Grant-In-Aid for the Scientific Research (C2) of the Ministry
of Education, Science, Sports and Culture in Japan (11640273).

\newpage

\appendix


\section{Background Quantities}\label{Appendix:A}

\subsection{Connection coefficients}
\Eq{
{\bar \Gamma}^a_{bc}={}^m\!\Gamma^a_{bc}(y),\
{\bar \Gamma}^a_{ij}=-r D^a r \gamma_{ij},\
{\bar \Gamma}^i_{aj}=\frac{D_a r}{r}\delta^i_j,\
{\bar \Gamma}^i_{jk}=\hat\Gamma^i_{jk}(x).
}

\subsection{Curvature tensors}
\Eqr{
&& {\bar R}^a{}_{bcd}={}^m\! R^a{}_{bcd},\\
&& {\bar R}^i{}_{ajb}=-\frac{D_aD_b r}{r}\delta^i_j,\\
&& 
{\bar R}^i{}_{jkl}
=[K-(Dr)^2](\delta^i_k\gamma_{jl}-\delta^i_l\gamma_{jk}).
}

\subsection{Ricci tensors}
\Eqr{
&& {\bar R}_{ab}={}^m\!R_{ab}-\frac{n}{r}D_aD_b r,\\
&& {\bar R}_{ai}=0,\\
&& {\bar R}^i_j=\left[-\frac{\square 
r}{r}+(n-1)\frac{K-(Dr)^2}{r^2}\right]\delta^i_j,\\
&& {\bar R}={}^m\!R -2n\frac{\square 
r}{r}+n(n-1)\frac{K-(Dr)^2}{r^2}.
}

\subsection{Einstein tensors}
\Eqr{
&& {\bar G}_{ab}=\,{}^m\!G_{ab}-\frac{n}{r}D_aD_b r 
-\left[\frac{n(n-1)}{2}\frac{K-(Dr)^2}{r^2}
-\frac{n}{r}\square r\right]g_{ab} \\
&& {\bar G}^i_j
=\left[-\frac{1}{2}{}^m\!R-\frac{(n-1)(n-2)}{2}\frac{K-(Dr)^2}
{r^2}+\frac{n-1}{r}\square r \right]\delta^i_j\\
&& {\bar G}_{ai}=0.
}


\section{Perturbations of the Ricci Tensors of the Bulk}
\label{Appendix:B}

In general the perturbation of the Ricci tensor is expressed in 
terms of $h_{MN}=\delta {\bar g}_{MN}$ as
\Eqr{
&2\delta {\bar R}_{MN} = & -{\bar\nabla}^L{\bar\nabla}_L 
h_{MN}-{\bar\nabla}_M{\bar \nabla}_N h
+{\bar\nabla}_M{\bar \nabla}_L h^L_N 
+ \bar\nabla_N\bar\nabla_L h^L_M \nonumber\\
&& +{\bar R}_{ML}h^L_N+ {\bar R}_{NL}h^L_M-2 {\bar 
R}_{MLNS}h^{LS},\\
&\delta {\bar R} =& -h_{MN}{\bar R}^{MN}+\bar\nabla^M\bar\nabla^N 
h_{MN}-\bar\nabla^M\bar\nabla_M h.
}
By decomposing the connection $\nabla$ into $D$ and $\hat D$ we 
obtain
\Eqr{
& 2\delta {\bar R}_{ab}= 
& -\square h_{ab}+D_aD_ch^c_b+D_bD_ch^c_a \nonumber\\
&& +n\frac{D^cr}{r}(-D_ch_{ab}+D_ah_{cb}+D_bh_{ca})\nonumber\\
&&+{}^m\!R^c_ah_{cb}+{}^m\!R^c_bh_{ca}-2\,{}^m\!R_{acbd} 
h^{cd}-\frac{1}{r^2}\hat\triangle h_{ab}\nonumber\\
&& +\frac{1}{r^2}(D_a\hat D^ih_{bi}+D_b\hat D^ih_{ai})
-\frac{D_br}{r^3}D_a h_{ij}\gamma^{ij}
-\frac{D_ar}{r^3}D_b h_{ij}\gamma^{ij}
\nonumber\\
&& +\frac{4}{r^4}D_arD_br h_{ij}\gamma^{ij}-D_aD_b h,
}
\Eqr{
& 2\delta {\bar R}_{ai}= 
& \hat D_iD_b h^b_a+\frac{n-2}{r}D^br \hat D_i h_{ab} \nonumber\\
&&-r\,\square\left(\frac{1}{r}h_{ai}\right)
-\frac{n}{r}D^br D_b h_{ai}
-D_ar D_b\left(\frac{1}{r}h^b_i\right)\nonumber\\
&& +\frac{n+1}{r}D^br D_ah_{bi}
+ rD_a D_b\left(\frac{1}{r}h^b_i\right) \nonumber\\
&& +\left[(n+1)\frac{(Dr)^2}{r^2}+(n-1)\frac{K-(Dr)^2}{r^2}
-\frac{\square r}{r}\right]h_{ia}
\nonumber\\
&& + \frac{1}{r^2}D^br D_ar h_{bi}
+(n+1)rD_a\left(\frac{1}{r^2}D^br\right)h_{bi} \nonumber\\
&& -\frac{n+2}{r}D_aD^br h_{ib}
+ {}^m\!R^b_a h_{bi}-\frac{1}{r^2}\hat\triangle h_{ai}
+\frac{1}{r^2}\hat D_i\hat D^j h_{aj}
\nonumber\\
&& +rD_a\left(\frac{1}{r^3}\hat D^jh_{ji}\right)
+\frac{1}{r^3}D_ar \hat D^j h_{ji}
-\frac{1}{r^3}D_ar \hat D_i h_{jk}\gamma^{jk} \nonumber\\
&& -rD_a\left(\frac{1}{r}\hat D_i h\right),
}
\Eqr{
&2\delta {\bar R}_{ij}= 
& \left[2rD^ar D_bh^b_a+2(n-1)D^arD^br h_{ab}+2rD^aD^br 
h_{ab}\right]\gamma_{ij}\nonumber\\
&& +r \hat D_i D_a\left(\frac{1}{r}h^a_j\right)
+r\hat D_jD_a\left(\frac{1}{r}h^a_i\right) \nonumber\\
&& +(n-1)\frac{D^ar}{r}(\hat D_ih_{aj}+\hat D_jh_{ai})
 +2\frac{D^ar}{r}\hat D^kh_{ka}\gamma_{ij}
\nonumber\\
&&-r^2\,\square \left(\frac{1}{r^2}h_{ij}\right)
-n\frac{D^ar}{r}D_ah_{ij}
+\frac{1}{r^2}(\hat D_i\hat D^k h_{kj}+\hat D_j\hat D^k h_{ki})
\nonumber\\
&&-\frac{1}{r^2}\hat\triangle h_{ij}
+2\left[(n-1)\frac{K}{r^2}+2\frac{(Dr)^2}{r^2}-\frac{\square r}{r}
\right]h_{ij} \nonumber\\
&&
-2(\gamma^{kl}h_{kl}\gamma_{ij}-h_{ij})\frac{K-(Dr)^2}{r^2}-2\frac{(D
r)^2}{r^2}\gamma_{ij}\gamma^{kl}h_{kl} \nonumber\\
&& -\hat D_i\hat D_j h- rD^arD_a h \gamma_{ij},
}
\Eqr{
&\delta {\bar R} = 
& D_aD_b h^{ab}+\frac{2n}{r}D^a rD^b h_{ab} \nonumber\\
&& +\left(-{}^m\!R^{ab}+\frac{2n}{r}D^aD^br 
+\frac{n(n-1)}{r^2}D^ar D^b r\right)h_{ab}
\nonumber\\
&& +\frac{2}{r^2}D_a\hat D^i h_i^a
+2(n-1)\frac{D^ar}{r^3}\hat D^i h_{ai}
\nonumber\\
&& +\frac{1}{r^4}\hat D^i \hat D^jh_{ij}
-\frac{D^ar}{r^3}D_a h_{ij}\gamma^{ij}
-\frac{1}{r^2}\left[(n-1)\frac{K}{r^2}-2\frac{(Dr)^2}{r^2}
\right]h_{ij}\gamma^{ij}
\nonumber\\
&&-\square h - n\frac{D^ar}{r}D_a h
-\frac{1}{r^2}\hat\triangle h
}
%


\section{Scalar Master Variable}
\label{Appendix:C}

In this appendix we show by a method different from the proof given in
\cite{Mukohyama.S2000A} that $F_{ab}$ and $F$ satisfying
\eqref{BulkPEq:vacuum1} and \eqref{BulkPEq:vacuum2} are written in
terms of the master variable $\Omega$ as in \eqref{MV:maxsymm1}and
\eqref{MV:maxsymm2} if the 2-dimensional orbit space with 
the metric $g_{ab}$ is a space $\N$ with a constant 
sectional curvature $\lambda$. 

First, let $W_{ab}$ be a symmetric, traceless, and divergenceless tensor
field on $\N$. Let $\xi^a$ be a (time-like) Killing vector, which
exists because $\N$ is maximally symmetric. If we put
$W_a=W_{ab}\xi^b$, from the divergenceless condition and the Killing
equation, we obtain
\Eq{
D_a W^a=W_{ab}D^a\xi^b=0.
\label{W:div1}}
In the same way, we obtain the conservation law for the combination
$W_{ab}\epsilon^{bc}\xi_c$ as  
\Eq{
D^a(W_{ab}\epsilon^{bc}\xi_c)=W^a_{b}\epsilon^{bc}D_{a}\xi_{c}
=-W_{ab}\epsilon^{bc}\epsilon^a{}_c
\left(\frac{1}{2}\epsilon^{ef}D_e\xi_f\right)
=W^c_c\left(\frac{1}{2}\epsilon^{ef}D_e\xi_f\right)=0.
\label{W:div2}}
Here, from the traceless condition, this vector is related to $W_a$
as
\Eq{
W_{ab}\epsilon^{bc}\xi_c=-\epsilon^{ab}W_{bc}\epsilon^{ce}\epsilon_{ef}\xi^f=-\epsilon_{ab}W^b.
}
Hence \eqref{W:div2} is written as
\Eq{
\epsilon^{ab}D_aW_b=0,
}
which implies that $W_a$ is written as a gradient of a function $W$:
\Eq{
W_a=D_aW.
\label{W:def}}
\eqref{W:div1} yields the Laplace equation 
\Eq{
\square W=0.
\label{W:Laplace}}

Since the vector defined by $\eta_a=\epsilon_{ab}\xi^b$ is 
orthogonal to $\xi^a$ and has the norm $\eta_a\eta^a=-\xi_a\xi^a$, 
the metric $g_{ab}$ is written as%
\Eq{
g_{ab}=\frac{1}{U}(- \xi_a\xi_b + \eta_a\eta_b),
}
where $U=-\xi_a\xi^a$. Utilizing this and the traceless condition, we obtain
\Eqr{
&W_{ab} & =W_{ac}\delta^c_a
= -\frac{1}{U}(W_a\xi_b+\epsilon_{ac}W^c\eta_b)
\nonumber\\
&& = - \frac{1}{U}(W_a\xi_b+W_b\xi_a-g_{ab}W_c\xi^c).
\label{Wab:GeneralSol}}
It is easily checked that the right-hand side of this equation is a
symmetric, traceless, and divergenceless tensor if \eqref{W:def} and
\eqref{W:Laplace} are satisfied.

In order to apply this formula to our problem, let us introduce the
traceless tensor $Z_{ab}$ as
\Eq{
r^{n-2}F_{ab}=Z_{ab}-(n-2)r^{n-2}Fg_{ab}.
}
This tensor is not divergenceless:
\Eq{
D_b Z^b_a=nD_a(r^{n-2}F).
}
In order to define a divergenceless tensor, let us introduce
a variable $\Omega$ as
\Eq{
2n r^{n-2}F=(\square +2\lambda)\Omega,
}
and define $W_{ab}$ as
\Eq{
W_{ab}=Z_{ab}-\left(D_aD_b\Omega-\frac{1}{2}\square\Omega g_{ab}\right).
}
It is easy to check that $W_{ab}$ is a symmetric, traceless, and 
divergenceless tensor if $\lambda$ is constant, hence it is written in
terms of a potential $W$ as in \eqref{Wab:GeneralSol}.

Here, note that in the definition of $\Omega$ there exists a freedom of
replacement $\Omega \tend \Omega+\phi$ where $\phi$ is a solution of
the hyperbolic equation
\Eq{
(\square +2\lambda)\phi=0.
}
By this replacement $W_{ab}$ changes as
\Eq{
W_{ab} \tend W'_{ab}=W_{ab}-(D_aD_b\phi+\lambda\phi g_{ab}).
}
Since $\phi$ is constrained by the hyperbolic equation, we can choose
the initial condition of $\phi$ and $\partial_t\phi$ on an initial
surface $t$=const so that $W'_{rr}=W'_{tr}=0$, where $t$ and $r$ are
the coordinates used in \eqref{BGsol:S}. This condition is written in
terms of the potential $W'$ for $W'_{ab}$ as $\partial_t W'=\partial_r
W'=0$. For any boundary condition on $W'$ that is linear and gives a
well-posed initial value problem, the solution satisfying this initial
condition is $W'$=const, which implies that $W'_{ab}=0$. Thus $F_{ab}$
and $F$ are expressed as in
\eqref{MV:maxsymm1} and \eqref{MV:maxsymm2}.


\section{General Solution of Equation \eqref{PDE}}
\label{Appendix:D}

In this appendix we give the general solution to
\eqref{PDE} on a two-dimensional maximally symmetric 
space. We work in the coordinates $(t,r)$ used in 
\eqref{BGsol:S}. Since the general solution for the case 
$\lambda=0$ is obviously given by $E=C_0+C_1t +C_2r$ with 
arbitrary constants $C_0\sim C_2$, we assume 
$\lambda\not=0$ below.

First, note that in the $(t,r)$ coordinates the non-vanishing
Christoffel symbols are given by
\Eq{
\Gamma^t_{tr}=\frac{U'}{2U},\ 
\Gamma^r_{tt}=\frac{1}{2}UU',\ 
\Gamma^r_{rr}=-\frac{U'}{2U}.
}
From this equation the $(tr)$-component of \eqref{PDE} is written as
\Eq{
0=D_tD_r E={U^{1/2}}\partial_r \left(U^{-1/2} \partial_t E\right),
}
which yields
\Eq{
E=f_1(t)U^{1/2}+f_2(r).
}
Inserting this expression into the $(tt)$-component of \eqref{PDE}, we obtain
\Eq{
0=(D_tD_t+\lambda g_{tt})E
=U^{1/2}\left[\ddot f_1-\left(\lambda U+\frac{(U')^2}{4}\right)f_1\right]
-U\left(\frac{1}{2}U'f_2'+\lambda f_2\right),
\label{PDE:tt}
}
where the overdot and the prime denote the differentiation with
respect to $t$ and $r$, respectively. Since $\lambda
U+(U')^2/4=\lambda K$ is constant, this equation is equivalent to the
following two ordinary differential equations: 
\Eqr{
&& \ddot f_1-\lambda K f_1=c,\\
&& -rf_2'+f_2=\frac{c}{\lambda U^{1/2}},
}
where $c$ is a separation constant. The general solution of the first
equation is given by
\Eq{
f_1(t)=\left\{\begin{array}{ll} 
	\frac{1}{2}ct^2+c_1t + c_0;& K=0\\
	-\frac{c}{\lambda K}+c_0 e^{\sqrt{\lambda K}t}
	+c_1 e^{-\sqrt{\lambda K}t}; & K\not=0.
	\end{array}\right.
}
On the other hand, the general solution for the equation for $f_2$ is
given by
\Eq{
f_2(r)=\left\{\begin{array}{ll} 
	c_2r + \frac{c}{2(-\lambda)^{3/2}r} ;& K=0\\
	c_2r+\frac{c}{\lambda K}U^{1/2}; & K\not=0.
	\end{array}\right.
}

Hence, after redefinitions of constants, the general solution
including the case $\lambda=0$ is expressed as
\Eq{
E=C_0 g_0(t,r) + C_1 g_1(t,r) + C_2 r,
}
where
\Eqr{
& g_0(r) & =\left\{\begin{array}{ll} 
	1; & \lambda=0, K\not=0\\
	\lambda^2 t^2r+ \frac{1}{r}; & \lambda\not=0, K=0\\
	e^{\sqrt{\lambda K}t}U^{1/2}; & \lambda K\not=0,
	\end{array}\right. \\
& g_1(r) & =\left\{\begin{array}{ll} 
	t; & \lambda K=0\\
	e^{-\sqrt{\lambda K}t}U^{1/2}; & \lambda K\not=0.
	\end{array}\right. 
}
It is easy to check that this satisfies the remaining $(rr)$-component
of \eqref{PDE}
\Eq{
0=(D_rD_r+\lambda g_{rr})E
=\left(\partial_r^2+\frac{U'}{2U}\partial_r 
+\frac{\lambda}{U}\right)E.
}
%


\section{Exceptional Modes for Scalar Perturbation with $K>0$}
\label{Appendix:E}

In this appendix we show that the gauge-equivalent classes of the
solutions to the perturbed Einstein equations are parametrized by a
finite number of parameters for the exceptional modes $k^2(k^2-nK)=0$
($K>0$) of the bulk scalar perturbation on a maximally symmetric
background.

First, let us consider the mode $k^2=0$. For this mode $\SHB_i$ and
$\SHB_{ij}$ vanish, and $f_a$ and $H_T$ are undefined. Further, the
gauge transformation is parametrized only by $T_a$. Hence, setting the
undefined variables to zero, $F$ and $F_{ab}$ are written as $F=H_L$
and $F_{ab}=f_{ab}$, which transform under the gauge transformation as
\Eq{
\bar\delta F=-\frac{D_ar}{r}T^a,\ 
\bar\delta F_{ab}=-D_a T_a - D_b T_a.
\label{GT:k=0}}

For the same reason, the equations 
\eqref{BulkPerturbationEq:scalar2} and 
\eqref{BulkPerturbationEq:scalar4}, or equivalently, 
\eqref{BulkPEq:vacuum1} and \eqref{BulkPEq:vacuum1}, do not exist 
for the mode $k^2=0$. However, we can recover these equations by
regarding them as the gauge-fixing conditions. Then the residual gauge
freedom is represented by $T_a$ satisfying the following two
conditions:
\Eqr{
&0 &=\bar\delta\left[\tilde F^a_a+2(n-2)\tilde F\right]
=-2D_a\tilde T^a,
\label{GC:k=0:1}\\
&0 &=\bar\delta\left[D_b\tilde F^b_a-2D_a\tilde F\right]
\nonumber\\
&& = -\square\tilde T_a+\frac{n-2}{r}Dr\cdot D\tilde T_a
+\frac{2}{r}D^br D_a\tilde T_b \nonumber\\
&& \quad -\left\{(n-2)\frac{K}{r^2}+(2n-1)\lambda\right\}\tilde T_a
-\frac{n}{r^2}D_arD_br\tilde T^b \nonumber\\
&& \quad +D_aD_b\tilde T^b+\frac{n-2}{r}D^ar D_b\tilde 
T^b,\label{GC:k=0:2}
}
where 
\Eq{
\tilde F=r^{n-2}F,\ 
\tilde F_{ab}=r^{n-2}F_{ab},\
\tilde T_a=r^{n-2}T_a.
}

Equation \eqref{GC:k=0:1} implies that $\tilde T^a$ is represented 
by a scalar function $T$ as
\Eq{
\tilde T^a=\epsilon^{ab}D_bT,
}
because the orbit space $\N$ is two dimensional. Inserting this 
expression into \eqref{GC:k=0:2}, we obtain
\Eq{
\epsilon^{ab}D_b\left[r^2\square T-nrDr\cdot DT+2(n-1)KT\right]=0.
}
Hence, by replacing $T$ by $T + $const, we obtain
\Eq{
r^2\square T-nrDr\cdot DT+2(n-1)KT=0.
\label{ResidualGauge}
}

Since \eqref{BulkPEq:vacuum1} and \eqref{BulkPEq:vacuum2} hold under
the above gauge conditions, any solution of the perturbed Einstein
equations is parametrized by $\Omega$ satisfying $(D_aD_b+\lambda
g_{ab})E(\Omega)=0$ as for the generic mode. Let the set of solutions
$\Omega$ to this equation be $\SS_\Omega$. Then we have an onto map
$\Phi_1$ from $\SS_\Omega$ to the space of solutions to the perturbed
Einstein equations. The kernel of this map is spanned by the solutions
of $(D_aD_b+\lambda g_{ab})\Omega=0$. On the other hand, $F$ and
$F_{ab}$ obtained by setting $F=\bar\delta F$ and $F_{ab}=\bar\delta
F_{ab}$ in \eqref{GT:k=0} with $T_a$ satisfying the above gauge-fixing
condition is also a solution to the perturbed Einstein equations
belonging to the trivial gauge-equivalent class. This correspondence
defines a map $\Phi_2$ from the space $\SS_G$ of solutions $T$ to
\eqref{ResidualGauge}. Then the set $\SS_\r{inv}$ of 
gauge-equivalence classes to the perturbed Einstein equations is 
represented as $\SS_\Sigma/\Phi_1^{-1}\Phi_2\SS_G$.

Here, note that $\SS_\Sigma/\ker \Phi_1$ is parametrized by the
solution to the equation $E(\Omega)=C_2r$, and hence by the initial data
$(\Omega,\dot\Omega)$ on an initial surface and the constant
$C_2$. Similarly, $\SS_G$ is parametrized by the initial data $(T,\dot
T)$ for \eqref{ResidualGauge}. Therefore, by representing the
condition $\Phi_1(\Omega)=\Phi_2(T)$ as a relation between these
initial data (and $C_2$), we can determine $\SS_\r{inv}$.

Now let us undertake this program. First, by redefining $-T$ as $T$,
the condition $\Phi_1(\Omega)=\Phi_2(T)$ is expressed as
\Eq{
(D_aD_b+\lambda g_{ab})\Omega =
\epsilon_{ac}D_bD^cT+\epsilon_{bc}D_aD^cT
-\frac{n-2}{r}\left(D_ar\epsilon_{bc}+D_br\epsilon_{ac}\right)D^cT
+\frac{2(n-1)}{r}\epsilon_{cd}D^crD^dT g_{ab}.
}
In the $(t,r)$-coordinates used in \eqref{BGsol:S}, with the help of
the equations for $\Omega$ and $T$, the trace and $(t,r)$-component of
this equation are written as
\Eqr{
&& U\Frac(\Omega/r)'+\frac{C_2}{nr}=\frac{2}{r}\dot T,\\
&& U^{1/2}\left(U^{-1/2}\dot\Omega\right)'
=2UT''+\left\{U'-2(n-1)\frac{U}{r}\right\}T'
+2(n-1)\frac{K}{r^2}T.
}
These equations have a solution for $(T,\dot T)$ when data
$(\Omega,\dot \Omega)$ is given.

On the other hand, the $(r,r)$-component is expressed as
\Eq{
U\Omega''+\frac{1}{2}U\Omega'+\lambda \Omega
=2U^{1/2}\left(U^{-1/2}\dot T\right)'+\frac{2}{r}\dot T,
}
and gives a constraint on $C_2$. In fact, inserting the expression for
$\dot T$ obtained from the trace, we obtain the condition $C_2=0$. 
This implies that the set $\Phi_1^{-1}\Phi_2\SS_G$ coincides with the set of
solutions to the homogeneous equation $E(\Omega)=0$. Thus $C_2$ is a
gauge-invariant and parametrizes the space $\SS_\r{inv}$.

Next we examine the mode $k^2=nK$. The argument is almost the same as
in the above case. Now the harmonic scalar and the harmonic vector are
non-trivial but the harmonic tensor $\SHB_{ij}$ vanishes. Hence only
$H_T$ is undefined, and $X_a$ is defined as $X_a=rf_a/k$. The gauge
transformation of $F$ and $F_{ab}$ are given by
\Eqr{
&&\bar\delta F=-\frac{r}{k}\left[Dr\cdot D\Frac(L/r)
+\frac{K}{r^2}L\right],\\
&&\bar\delta F_{ab}=-\frac{1}{k}
\left[D_a\left\{r^2D_b\Frac(L/r)\right\}
+D_b\left\{r^2D_a\Frac(L/r)\right\}\right].
}

In the present case only the equation \eqref{BulkPEq:vacuum1} is
lacking. Hence we regard this as the gauge-fixing condition. Then the
residual gauge freedom is parametrized by $L$ satisfying the wave
equation
\Eq{
\square \tilde L-\frac{n}{r}Dr\cdot\tilde L
+\left(n\lambda+2(n-1)\frac{K}{r^2}\right)\tilde L=0,
}
where $\tilde L=r^{n-1}L$. After the redefinition $-2\tilde L/k
\tend \tilde L$, the condition $\Phi_1(\Omega)=\Phi_3(L)$ is 
represented as
\Eqr{
&(D_aD_b+\lambda g_{ab})\Omega=
& D_aD_b\tilde L-\frac{n-1}{r}\left(D_ar D_b\tilde L
+D_br D_a\tilde L\right)+\frac{n(n-1)}{r^2}D^arD^br\tilde L
\nonumber\\
&&+\left[\frac{n-1}{r}Dr\cdot D\tilde L
+\left\{n^2\lambda-(n-1)^2\frac{K}{r^2}\right\}\tilde L\right]g_{ab}.
}
Here $\Phi_3$ represents the map from the space $\SS_L$ of solutions 
$L$ to the set of solutions to the perturbed Einstein equations. 

The trace and the $(t,r)$-component of this equation are written as
\Eqr{
&& U\Omega'+\lambda r\Omega+\frac{1}{nr}\left(C_0g_0+C_1g_1\right)
=U\tilde L'+\left[n\lambda r-(n-1)\frac{K}{r}\right]\tilde L,\\
&& U^{1/2}\left(U^{1/2}\dot\Omega\right)'
=U^{1/2}r^{n-1}\left(\frac{\dot{\tilde L}}{U^{1/2}r^{n-1}}\right)',
}
which have a solution for $(L,\dot L)$ for any data $(\Omega,\dot 
\Omega)$. On the other hand, the $(r,r)$-component
\Eq{
U\Omega''-\lambda r\Omega'+\lambda\Omega
=U\tilde L''-\left(\lambda r+\frac{n-1}{r}U\right)\tilde 
L'+\left[(n-1)\frac{K}{r^2}+n\lambda\right]\tilde L,
}
gives the constraint $C_0=C_1=0$. Thus $\Phi_1^{-1}\Phi_3\SS_T$
coincides with the space of solutions to the homogeneous equation
$E(\Omega)=0$, and the space $\SS_\r{inv}$ of the gauge-equivalence
classes of solutions is parametrized by the two gauge-invariant
constants $C_0$ and $C_1$.



\end{document}